\documentclass[aps,prd,preprint,showkeys,groupedaddress]{revtex4-1}
\usepackage[dvipdfmx]{graphicx,color}
\usepackage{amsmath,bm}
\usepackage{amssymb}
\usepackage{multirow}

\begin{document}

\preprint{KEK-TH-2391 and J-PARK-TH-0268}

\title{Physical symmetries and gauge choices in the Landau problem}


%

\author{Masashi Wakamatsu$^{1}$}
\email{wakamatu@post.kek.jp}
\author{Akihisa Hayashi$^{2}$} 
\email{hayashia@u-fukui.ac.jp}
\affiliation{$^{1}$KEK Theory Center, Institute of Particle and Nuclear Studies, High Energy Accelerator
Research Organization (KEK), 1-1, Oho, Tsukuba, Ibaraki 305-0801, Japan}
\affiliation{$^{2}$Department of Applied Physics, University of Fukui,
Bunkyo 3-9-1, Fukui-City, Fukui 910-8507, Japan}
%


\date{\today}

\begin{abstract}
Due to a special nature of the Landau problem, in which the magnetic
field is uniformly spreading over the whole two-dimensional plane,
there necessarily exist three conserved quantities, i.e. 
two conserved momenta and one conserved orbital angular momentum
for the electron, independently of the choice of the gauge potential.
Accordingly, the quantum eigen-functions of the Landau problem can
be obtained by diagonalizing the Landau Hamiltonian together with one of the 
above three conserved operators with the result that the quantum mechanical 
eigen-functions of the Landau problem can be written down for arbitrary 
gauge potential. 
The purpose of the present paper is to clarify the meaning of gauge choice
in the Landau problem based on this gauge-potential-independent formulation, 
with a particular intention of unraveling the physical significance of the concept of 
gauge-invariant-extension of the canonical orbital angular momentum 
advocated in recent literature on the nucleon spin decomposition problem.
At the end, our analysis is shown to disclose a physically vacuous side face
of the gauge symmetry. 
\end{abstract}


\maketitle
 

\section{Introduction}\label{sec1}

For a quantum mechanical formulation of the Landau problem, one must necessarily 
introduce gauge potential instead of magnetic field 
itself \cite{Landau1930},\cite{Landau-Lifschitz}. 
The problem here is that the gauge potential,
which reproduces the given magnetic field, is not unique at all.
Practically useful choices of gauge are known to be the two Landau gauges and 
the symmetric gauge. It is widely believed that the choice of a particular gauge 
is inseparably connected with a specific symmetry of the corresponding electron 
eigen-functions \cite{Vallentine1998}\nocite{VGW2006}\nocite{Goerbig2009}
\nocite{Tong2016}-\cite{Murayama}.
In fact, the choice of one of the two Landau gauges naturally leads to the
Hermite-type eigen-functions, which respects the translational symmetry
along the $x$-direction or $y$-direction. On the other hand, the choice of
symmetric gauge leads to the Laguerre-type eigen-functions, which respects
the rotational symmetry around the coordinate origin.
However, although it is not so familiar, the connection between the choice of 
gauge potential and the symmetry of the electron eigen-states is not an 
absolute demand of the quantum theory.
The reason is because, in the Landau problem, there exist two conserved momenta
and one conserved orbital angular momentum (OAM) of the electron 
irrespectively of the choice of the gauge potential. 
(In some previous literature, these conserved quantities were called the
pseudo momenta and the pseudo OAM \cite{Yoshioka2002}\nocite{Konstantinou2016}
-\cite{Konstantinou2017}.)
In persuit of the meaning of gauge choice in the Landau problem, the authors
of the paper \cite{WKZ2018} solved the Landau problem 
based on the gauge-invariant (but path-dependent) formulation of 
the quantum electrodynamics proposed by DeWitt many years ago \cite{DeWitt1962}.
In this theoretical framework, though the solutions of the Landau problem
can be written down without specifying the gauge potential, they instead
depend on the choice of path of the nonlocal gauge link also called the Wilson line.
Choosing rectangular paths connecting the coordinate origin and the position
of the relevant fields in the rectangular coordinate representation, it
leads to a class of solutions, which may be regarded as a gauge-invariant
extension based on either of the two Landau gauges. On the other hand,
choosing a straight-line path connecting the coordinate origin and the
position of the relevant fields in the 2-dimensional spherical coordinate system, 
it leads to a class of solutions,
which can be regarded as a gauge-invariant extension based on the symmetric
gauge. Using these two classes of solutions, they analysed the expectation values 
of the six operators \cite{WKZ2018}. Those are the canonical momentum, 
pseudo momentum, and mechanical momentum operators of the electron
as well as the canonical OAM, pseudo OAM, and the mechanical OAM operators. 
This analysis clarified the nature of
the pseudo momentum as a gauge-covariant extension of the canonical
momentum based on the Landau gauge, and also the nature of the
pseudo OAM as a gauge-covariant extension of the canonical OAM based
on the symmetric gauge. Unfortunately, the achievement of their analysis
is still half way, in the sense that they did not fully elucidate the relations between 
the two classes of solutions, i.e. the gauge-potential-independent extension 
based on the two Landau gauge eigen-states and that based on the symmetric 
gauge eigen-states. 
Particularly interesting questions to be asked here would be the following.
First, do the expectation values of the mechanical momentum and that of the 
mechanical OAM in fact coincide in these two different classes of 
eigen-functions ? 
Since it is generally believed that both of those are manifestly gauge-invariant 
quantities, it seems natural to expect that the answer is ``Yes''.
If it is really so, what would be an answer to the same question asked for the 
pseudo momenta and the pseudo OAM ? 
The second question is especially interesting, because the pseudo momentum and 
pseudo OAM operators are known to transform {\it covariantly} 
under an arbitrary gauge transformation just like the mechanical momentum and
the mechanical OAM do. Aiming at answering the above nontrivial questions as 
transparently as possible, we propose in the present paper a 
gauge-potential-independent formulation of the Landau problem without using 
nonlocal gauge link. Instead, the treatment is based on more familiar 
unitary transformation method in the standard non-relativistic quantum mechanics. 
In this formulation, the eigen-functions of the Landau problem will be obtained by 
diagonalizing the Landau Hamiltonian together with one of the above three conserved 
quantities for arbitrary choice of the gauge potential, which causes that the quantum
mechanical eigen-functions of the Landau problem can be expressed for arbitrary
choice of the gauge potential. 
As a nontrivial byproduct, we will be able to show that this elementary formulation 
of the Landau problem helps us to unravel the physically insubstantial nature of 
the concept of gauge-invariant-extension of the canonical OAM 
advocated in recent literature on the nucleon spin decomposition 
problem \cite{Chen2008}\nocite{Chen2009}\nocite{Hatta2011}\nocite{Lorce2013}
-\cite{Review_LL2014}. (For reviews on the nucleon spin decomposition problem,
see \cite{Review_LL2014},\cite{Review_Waka2014}.)

The paper is organized as follows.
In Sect.\ref{sec2}, we explain the reason why there exist three conserved 
quantities in the Landau problem, i.e. two conserved momenta
and one conserved OAM, independently of the
choice of the gauge potential. Utilizing this fact, we give a
gauge-potential-independet formulation of the Landau problem,
in which the eigen-functions of the Landau problem 
are written down for arbitrary gauge potentials.
We show that, in this formulation, the eigen-functions of the Landau problem
are devided into three classes, which we would call the 
gauge-potential-independent extension based on the 1st Landau gauge 
eigen-states, that based on the 2nd Landau gauge eigen-states, and that 
based on the symmetric 
gauge eigen-states.
Next, in Sect.\ref{sec3}. we evaluate the matrix elements of three kinds of
momentum and OAM operators, i.e. the canonical ones, the mechanical ones,
and the conserved ones, between the Landau eigen-states belonging
to different classes mentioned above, with a particular intention of verifying
gauge-independence or gauge-dependence of the matrix
elements of the two types of quantities, i.e. the mechanical momentum and the
mechanical OAM, and the conserved momentum and the conserved OAM.
In Sect.\ref{sec4}, we try to make clear the physical meaning of several
findings reported in the previous sections. 
In particular, we argue that the pseudo
momenta as well as the pseudo OAM are not gauge-invariant physical
quantities in the standard or physical sense. 
Finally, in Sect.\ref{sec5}, we summarize what we can learn from the 
present investigation, especially on about highly delicate nature of the gauge 
symmetry concept in physics.

\section{Symmetries of the Landau system and conserved quantities
for arbitrary choice of gauge potential}\label{sec2}

The Landau problem deals with the motion of an electron in a uniform magnetic field
perpendicular to the $xy$-plane given as $\bm{B} (\bm{r}) = B \,\bm{e}_z$
with $B$ being a constant.
Denoting the mass and the charge of the electron as $m_e$ and
$- \,e$ with $e > 0$, the classical equations of motion of the electron
are given by
\begin{eqnarray}
 m_e \,\ddot{x} \ &=& \ - \,e \,B \,\dot{y}, \\
 m_e \,\ddot{y} \ &=& \ + \,e \,B \,\dot{x},
\end{eqnarray}
where $\dot{x} = \frac{d x}{d t}, \dot{y} = \frac{d y}{d t}$ and
$\ddot{x} = \frac{d^2 x}{d t^2}, \ddot{y} = \frac{d^2 y}{d t^2}$.
The familiar {\it mechanical} (or {\it kinetic}) momenta and the 
{\it mechanical} (or {\it kinetic}) orbital 
angular momentum (OAM) of the electron are respectively 
represented as
\begin{eqnarray}
 p^{mech}_x \ &=& \ m_e \,\dot{x}, \\
 p^{mech}_y \ &=& \ m_e \,\dot{y}, \\
 L^{mech}_z \ &=& \ m_e \,\left( x \,\dot{y} \ - \ y \,\dot{x} \right).
\end{eqnarray}
Although it is not very familiar, by using
the above equations of motion, it is easy to show that the
following two momenta and one OAM are constants of motion, i.e.
$\frac{d}{d t} \,p^{cons}_x = 0$, $\frac{d}{d t} \,p^{cons}_y = 0$,
$\frac{d}{d t} \,L^{cons}_z = 0$ :
\begin{eqnarray}
 p^{cons}_x \ &\equiv& \ p^{mech}_x \ + \ e \,B \,y \ = \
 m_e \,\dot{x} \ + \ e \,B \,y , \\
 p^{cons}_y \ &\equiv& \ p^{mech}_y \ - \ e \,B \,x \ = \
 m_e \,\dot{y} \ - \ e \,B \,x , \\
 L^{cons}_z \ &\equiv& \ L^{mech}_z \ - \ \frac{1}{2} \,e \,B \,r^2
 \ = \ m_e \,\left( x \,\dot{y} \ - \ y \,\dot{x} \right) \ - \ 
 \frac{1}{2} \,e \,B \,r^2 .
\end{eqnarray}
The existence of these three conserved quantities is inseparably connected
with a special nature of the Landau 
problem \cite{Haugset1993},\cite{Yoshioka2002}\nocite{Konstantinou2016}-
\cite{Konstantinou2017}. That is, in the setting of the
Landau problem, the magnetic field is uniformly spreading over the whole 
2-dimensional plane, which means that there is no special point in this plane. 
One can choose any point as a coordinate origin
for describing the motion of the electron. In other words, the Landau system
has a translational invariance along an arbitrary direction in the 2-dimensional plane, 
which dictates the conservation of the two momenta $p^{cons}_x$ and
$p^{cons}_y$ along the two orthogonal directions.
Next, even after the coordinate origin is suitably fixed, there still
remains a freedom to choose the direction of two perpendicular axes.
This is equivalent to saying that the Landau system has a rotational symmetry
(or the axial symmetry) around the coordinate origin. Undoubtedly,
the Noether current corresponding this symmetry is the conserved OAM $L^{cons}_z$.  
Since the classical equations of motion can be written down without introducing
gauge-variant potential, the conservations of the above three quantities
has little to do with the problem of gauge choice in the Landau problem. 
We recall that, in some of the previous literature, these conserved quantities were 
called the {\it pseudo} momenta and {\it pseudo} OAM \cite{Yoshioka2002}
\nocite{Konstantinou2016}\nocite{Konstantinou2017}-\cite{WKZ2018}.
In the present paper, we shall call them the {\it conserved} momenta and
{\it conserved} OAM to emphasize that they are conserved quantities 
irrespectively of the choice of gauge potential in the Landau problem.
Complexity occurs, since the quantum mechanical formulation of the Landau 
problem needs to introduce the vector potential, which is necessarily gauge-dependent, 
and there is a wide-spread misbelief that the conservation of the 
above-mentioned quantities are inseparably connected with the choice of 
gauge or gauge potential. 
To understand this delicate issue of gauge choice in the Landau
problem in quantum mechanics, we think it instructive to first reformulate the
classical mechanics of the Landau electron within the Lagrangian formulation.

\subsection{Lagrangian formulation of the Landau problem}

\vspace{2mm}
Note that, in order to write down the Lagrangian of the Landau system,
we must introduce (gauge-variant) vector potential $\bm{A}$.
As is widely-known, there are three popular choices of gauge or gauge
potential in the Landau problem. They are 
the 1st Landau gauge $\bm{A}^{(L_1)}$, the 2nd Landau gauge
$\bm{A}^{(L_2)}$, and the symmetric gauge $\bm{A}^{(S)}$, specified by
\begin{eqnarray}
 \bm{A}^{(L_1)} (\bm{r}) \ &=& \ B \,\left( - \,y, 0 \right), \\
 \bm{A}^{(L_2)} (\bm{r}) \ &=& \ B \,\left( 0, x \right), \\
 \bm{A}^{(S)} (\bm{r}) \ &=& \ \frac{1}{2} \,B \,\left( - \,y, x \right) .
\end{eqnarray}
First, we consider the symmetric gauge. In this gauge, the Lagrangian
of the Landau system is given by
\begin{eqnarray}
 L (\bm{A}^{(S)} ) \ &=& \ \frac{1}{2} \,m_e \,\bm{v}^2 \ - \ 
 e \,\bm{v} \cdot \bm{A}^{(S)} \nonumber \\
 &=& \ \frac{1}{2} \,m_e \,\bm{v}^2 \ + \ \frac{1}{2} \,e \,B \,
 \left( \dot{x} \,y - \dot{y} \,x \right) \nonumber \\
 &=& \frac{1}{2} \,m_e \left( \dot{r}^2 \ + \ r^2 \,\dot{\phi}^2  \right)
 \ - \ \frac{1}{2} \,e \,B \,r^2 \,\dot{\phi} .
\end{eqnarray}
Here $r$ and $\phi$ are the radial and angular coordinates in the
2-dimensional polar coordinate system.
Since $L (\bm{A}^{(S)})$ does not depend on the {\it cyclic} coordinate $\phi$,
the corresponding canonical conjugate momentum defined by 
\begin{eqnarray}
 p_\phi (\bm{A}^{(S)}) \ &\equiv& \ 
 \frac{\partial L (\bm{A}^{(S)})}{\partial \dot{\phi}}
 \ = \ m_e \,r^2 \,\dot{\phi} \ - \ \frac{1}{2} \,e \,B \,r^2
 \ = \ L^{mech}_z \ - \ \frac{1}{2} \,e \,B \,r^2. \hspace{6mm}
\end{eqnarray}
should be conserved, i.e. $\frac{d}{d t} \,p_\phi (\bm{A}^{(S)}) = 0$.
Note that this just coincides with $L^{cons}_z$ given by (8), i.e. one
confirms that
\begin{equation}
 p_\phi (\bm{A}^{(S)}) \ = \ L^{cons}_z (\bm{A}^{(S)}) .
\end{equation}
Next, in the rectangular coordinate system, $L (\bm{A}^{(S)})$ {\it does} depend 
on $x$, but its partial derivative with respect to $x$ turns out to be a {\it total 
derivative} on {\it time} as
\begin{equation}
 \frac{\partial L (\bm{A}^{(S)})}{\partial x} \ = \ - \,\frac{1}{2} \,e \,B \,\dot{y} .
\end{equation}
This means that there must be a corresponding conserved quantity given by
\begin{equation}
 p_x (\bm{A}^{(S)}) \ + \ \frac{1}{2} \,e \,B \,y ,
\end{equation}
where $p_x (\bm{A}^{(S)})$ is the canonical conjugate momentum with
respect to $x$ defined by
\begin{equation}
 p_x (\bm{A}^{(S)}) \ \equiv \ \frac{\partial L (\bm{A}^{(S)})}{\partial \dot{x}} .
\end{equation}
Then, the conserved quantity indicated by (16) becomes
\begin{equation}
 p_x (\bm{A}^{(S)}) \ + \ \frac{1}{2} \,e \,B \,y
 \ = \ m_e \,\dot{x} \ + \ \frac{1}{2} \,e \,B \,y
 \ + \ \frac{1}{2} \,e \,B \,y \ = \ m_e \,\dot{x} \ + \ e \,B \,y .
\end{equation}
This just coincides with the conserved momentum $p^{cons}_x$ defined by (6), i.e.
we find that
\begin{equation}
 p_x (\bm{A}^{(S)}) \ + \ e \,B \,y \ = \ p^{cons}_x  .
\end{equation}
A lesson learned from the above analysis is therefore the following.
Although the conservation of $p^{cons}_x$ in the symmetric gauge is 
somewhat difficult to see as compared with $L^{cons}_z$, the former is 
nevertheless a conserved quantity even in the symmetric gauge.

Since the two Landau gauges are of similar character, we consider
below only the 1st Landau gauge specified by $\bm{A}^{(L_1)} = B \,(\,- \,y, 0)$.
The Lagrangian in this gauge is given as
\begin{eqnarray}
 L (\bm{A}^{(L_1)}) \ &=& \ \frac{1}{2} \,m_e \,\bm{v}^2 \ - \ 
 e \,\bm{v} \cdot \bm{A}^{(L_1)} 
 \ = \ \frac{1}{2} \,m_e \,\bm{v}^2 \ + \ e \,B \,\dot{x} \,y .
\end{eqnarray}
First, because $L (\bm{A}^{(L_1)})$ does not depend on the {\it cyclic} coordinate
$x$, the corresponding canonical momentum defined by: 
\begin{eqnarray}
 p_x (\bm{A}^{(L_1)}) \ &\equiv& \ \frac{\partial L (\bm{A}^{(L_1)})}{\partial \dot{x}}
 \ = \ m_e \,\dot{x} \ + \ e \,B \,y \ = \ p^{mech}_x \ + \ e \,B \,y. 
\end{eqnarray}
must be conserved, i.e. $\frac{d}{d t} \,p_x (\bm{A}^{(L_1)}) = 0$.
In fact, this just coincides with the conserved momentum $p^{cons}_x$ given 
by (6), i.e. we see that
\begin{equation}
 p_x (\bm{A}^{(L_1)}) \ = \ p^{cons}_x (\bm{A}^{(L_1)}).
\end{equation}
In contrast, in the polar coordinate, $L (\bm{A}^{(L_1)})$ {\it does} depend on
$\phi$. In fact, we have
\begin{eqnarray}
 L (\bm{A}^{(L_1)}) &=& \frac{1}{2} \,m_e \,( \dot{r}^2 \ + \ r^2 \,\dot{\phi}^2)
 \ + \ e \,B \,( \dot{r} \,\cos \phi - r \,\dot{\phi} \,\sin \phi) \,r \,\sin \phi
 \nonumber \\
 &=& \frac{1}{2} \,m_e \,( \dot{r}^2 \ + \ r^2 \,\dot{\phi}^2 ) \ + \ e \,B \,
 ( r \,\dot{r} \,\cos \phi \,\sin \phi - r^2 \,\dot{\phi} \,\sin^2 \phi) .
 \hspace{6mm}
\end{eqnarray}
Nonetheless, its derivative on $\phi$ turns out to be a {\it total derivative} 
with respect to {\it time} as
\begin{eqnarray}
 \frac{\partial L (\bm{A}^{(L_1)})}{\partial \phi} &=& e \,B \,r \,\dot{r} \,
  (\,- \,\sin^2 \phi \ + \ \cos^2 \phi ) \, - \, 
  e \,B \,r^2 \,\dot{\phi} \,\,2 \,\sin \phi \,\cos \phi \hspace{6mm} 
  \nonumber \\
  &=& \frac{d}{d t} \,\,\frac{1}{2} \,e \,B \,r^2 \,
  ( \cos^2 \phi \ - \ \sin^2 \phi ) .
\end{eqnarray}
This indicates that the following quantity should be conserved
\begin{equation}
 p_\phi (\bm{A}^{(L_1)}) \ - \ \frac{1}{2} \,e \,B \,r^2 \,
 ( \,\cos^2 \phi \ - \ \sin^2 \phi ),
\end{equation}
where $p_\phi (\bm{A}^{(L_1)})$ is the canonical conjugate momentum
with respect to $\phi$ in the $L_1$ gauge, defined by
\begin{equation}
 p_\phi (\bm{A}^{(L_1)}) \ \equiv \ 
 \frac{\partial L (\bm{A}^{(L_1)})}{\partial \dot{\phi}}.
\end{equation}
Note that the conserved quantity indicated by (25) can be rewritten as
\begin{eqnarray}
 &\,& p_\phi (\bm{A}^{(L_1)}) \ - \ \frac{1}{2} \,e \,B \,r^2 \,
 ( \,\cos^2 \phi \ - \ \sin^2 \phi ) \nonumber \\
 \ &=& \ m_e \,r^2 \,\dot{\phi} \ - \ e \,B \,r^2 \,\sin^2 \phi \ - \ 
 \frac{1}{2} \,e \,B \,r^2 \,( \cos^2 \phi - \sin^2 \phi) \hspace{6mm}
 \nonumber \\
 \ &=& \ m_e \,r^2 \,\dot{\phi} \ - \ \frac{1}{2} \,e \,B \,r^2. \ \ \ \ 
 \end{eqnarray}
This precisely coincides with the conserved angular momentum
$L^{cons}_z$ in the $L_1$-gauge, i.e. we find that
\begin{equation}
 p_\phi (\bm{A}^{(L_1)}) \ - \ \frac{1}{2} \,e \,B \,r^2 \,
 ( \,\cos^2 \phi \ - \ \sin^2 \phi ) \ = \ L^{cons}_z (\bm{A}^{(L_1)}).
\end{equation}
Thus, although the conservation of $L^{cons}_z$ is somewhat difficult to
see in the $L_1$ gauge, it nevertheless is a conserved quantity just 
like $p^{cons}_x$.

From the above considerations, it is clear that the following two
quantities are conserved in both of the symmetric gauge and the 1st
Landau gauge and that this fact can be extended to arbitrary gauge 
field configuration $\bm{A}$ : 
\begin{eqnarray}
 p^{cons}_x \ &=& \ p^{mech}_x (\bm{A}) \ + \ e \,B \,y , \label{Eq:p_cons_x} \\
 L^{cons}_z \ &=& \ L^{mech}_z (\bm{A}) \ - \ \frac{1}{2} \,e \,B \,r^2 .
 \label{Eq:L_cons_z}
\end{eqnarray}
(Although we do not repeat a similar analysis, it is evident
that the quantity $p^{cons}_y = p^{mech}_y (\bm{A}) - e \,B \,x$ is also
a conserved quantity independently of the choice of $\bm{A}$.)

At this point, we recommend readers to keep in mind the fact that, since the 
second terms in the r.h.s. of (\ref{Eq:p_cons_x}) and (\ref{Eq:L_cons_z}) are 
intact under a gauge transformations, 
the transformation property of $p^{cons}_x$ and $L^{cons}_z$ under an arbitrary 
gauge transformation is {\it exactly the same} as $p^{mech}_x$ and $L^{cons}_z$.
Namely, $p^{cons}_x$ and $L^{cons}_z$ transform {\it covariantly} 
under gauge transformation just like $p^{mech}_x$ and $L^{mech}_z$ do.
As we shall see in the rest of the paper, what is meant by this fact provides
with us a highly nontrivial question, the pursuit of whose answer leads us to
unexpectedly deep insight into delicate nature of the gauge symmetry 
concept.

\subsection{Hamilton formulation and quantum mechanics}

\vspace{2mm}
When going to quantum mechanics in coordinate representation, the
momentum operator is replaced by a derivative operator as 
$\hat{\bm{p}} \rightarrow - \,i \,\nabla$. 
(Throughout the paper, we use the natural unit, $\hbar = c = 1$.)
The Hamiltonian of the
Landau problem is then represented as
\begin{equation}
 \hat{H} (\bm{A}) \ = \ \frac{1}{2 \,m_e} \,\left( - \,i \nabla \ + \ 
 e \,\bm{A} \right)^2.
\end{equation}
The conserved momentum and the conserved OAM given by 
(\ref{Eq:p_cons_x}) and (\ref{Eq:L_cons_z})
also become quantum operators as
\begin{eqnarray}
 \hat{p}^{cons}_x \ &=& \ - \,i \,\frac{\partial}{\partial x} \ + \ 
 e \,A_x \ + \ e \,B \,y , \\
 \hat{L}^{cons}_z \ &=& \ - \,i \,\frac{\partial}{\partial \phi} \ + \ 
 e \,r \,A_\phi \ - \ \frac{1}{2} \,e \,B \,r^2 ,
\end{eqnarray}
where $A_x$ represents the $x$-component of the vector
potential $\bm{A}$, while $A_{\phi}$ stands for its azimuthal 
component. 
The classical conservations of these quantities
can be translated into the commutation relations beween the
corresponding quantum operators and the Landau Hamiltonian as
\begin{eqnarray}
 \,[\,\hat{p}^{cons}_x, \,\hat{H} (\bm{A}) \,] &=& 0, \\
 \,[\,\hat{L}^{cons}_z, \,\hat{H} (\bm{A}) \,] &=& 0.
\end{eqnarray}
In quantum mechanics, however, these two conserved quantities
$\hat{p}^{cons}_x$ and $\hat{L}^{cons}_z$ do not commute with
each other,
\begin{equation}
 [\, \hat{p}^{cons}_x, \,\hat{L}^{cons}_z \,] \ \neq \ 0 .
\end{equation}
This means that one is forced to choose either of these 
two operators as an operator which will be simultaneously diagonalized 
with the Landau Hamiltonian.

If one selects the symmetric gauge potential, it is conventional to choose
$\hat{L}^{cons} (\bm{A}^{(S)})$ as an operator to be diagonalized with the 
Landau Hamiltonian, so that the eigen-states 
$\vert \,\Psi^{(S)}_{n,m} \rangle$ in the symmetric gauge
is customarily taken as the simultaneous eigen-functions of 
$\hat{L}^{cons}_z (\bm{A}^{(S)})$
and $\hat{H} (\bm{A}^{(S)})$ as \cite{WKZ2018},\cite{Haugset1993}
\begin{eqnarray}
 \hat{L}^{cons}_z (\bm{A}^{(S)}) \,\vert \,\Psi^{(S)}_{n,m} \rangle \ &=& \ 
 m \, \vert \,\Psi^{(S)}_{n,m} \rangle , \label{Eq:Eigen_L_cons_z_Psi_S_n_m} 
 \label{Eq:eigen_L_cons} \\
 \hat{H} (\bm{A}^{(S)}) \,\vert \,\Psi^{(S)}_{n,m} \rangle \ &=& \  
 E_n \,\vert \,\Psi^{(S)}_{n,m} \rangle , \label{Eq:eigen_H_Landau}
\end{eqnarray}
where $E_n = (2 \,n + 1) \,\omega_L$ with $\omega_L = e \,B / (2 \,m_e)$
being the so-called Larmor frequency. 
In fact, in the symmetric gauge, $\hat{L}^{cons}_z (\bm{A})$ reduces to
\begin{equation}
 \hat{L}^{cons}_z (\bm{A}^{(S)}) \ \equiv \ - \,i \,\frac{\partial}{\partial \phi}
  \ + \ e \,r \,A^{(S)}_\phi \ - \ \frac{1}{2} \,e \,B \,r^2 \ = \ 
  - \,i \,\frac{\partial}{\partial \phi},
\end{equation}
which is nothing but the polar-coordinate representation of the
ordinary canonical OAM operator 
$\hat{L}^{can}_z = - \,i \,(\bm{r} \times \nabla)_z$. 
Accordingly, the eigen-functions of (\ref{Eq:eigen_L_cons}) and 
(\ref{Eq:eigen_H_Landau}) are given as
\begin{equation}
 \Psi^{(S)}_{n,m} (x, y) \ = \ \frac{e^{\,i \,m \,\phi}}{\sqrt{2 \,\pi}} \,\,
 N_{n,m} \,\,e^{\,- \,\frac{\xi}{2}} \,\,\xi^{\,\vert m \vert \,/\,2} \,\,
 L^{\vert m \vert}_{n - \frac{\vert m \vert + m}{2}} (\xi) ,
\end{equation}
where $\xi = r^2 \,/\,(2 \,l^2_B)$ with $l^2_B = 1 \,/\,(e \,B)$, while
\begin{equation}
 N_{n,m} \ = \ \frac{1}{l_B} \,\,
 \sqrt{\frac{\left( n \ - \ \frac{\vert m \vert + m}{2} \right) \, !}
 {\left( n \ + \ \frac{\vert m \vert - m}{2} \right) \,!}} , 
\end{equation}
is the normalization constant, and $L^{\alpha}_{n} (\xi)$ is the familiar 
associated Laguerre polynomials. 

Although somewhat unconventional, however, even with the choice
of the symmetric gauge potential $\bm{A}^{(S)}$,
one {\it can} make a choice in which $\hat{p}^{cons}_x (\bm{A}^{(S)})$ and 
$\hat{H} (\bm{A}^{(S)})$ are simultaneously diagonalized.
(This is so because $\hat{p}^{cons}_x$ is also a conserved quantity
irrespectively of a choice of the gauge potential.)
This amounts to choosing the basis vectors 
$\vert \,\Psi^{(S)}_{n, k_x} \rangle$ defined by the simultaneous
eigen-equations : 
\begin{eqnarray}
 \hat{p}^{cons}_x (\bm{A}^{(S)}) \,\vert \,\Psi^{(S)}_{n, k_x} \rangle \ &=& \ 
 k_x \,\vert \,\Psi^{(S)}_{n, k_x} \rangle, \\
 \hat{H} (\bm{A}^{(S)}) \,\vert \,\Psi^{(S)}_{n, k_x} \rangle \ &=& \ 
 E_n \,\vert \,\Psi^{(S)}_{n, k_x} \rangle .
\end{eqnarray}
Here, we have assumed that the eigen-energies of the Landau Hamiltonian
depends only on the Landau quantum number $n$, which is in fact the case 
as $E_n = \left( 2 \,n + 1\right) \,\omega_L$.

\vspace{3mm}
\noindent
Next, if one chooses to work by selecting the 1st Landau gauge 
potential $\bm{A}^{(L_1)}$, the relevant conserved quantities take the form : 
\begin{eqnarray}
 \hat{p}^{cons}_x (\bm{A}^{(L_1)}) \ & = & \ - \,i \,\frac{\partial}{\partial x} , \\
 \hat{L}^{cons}_z (\bm{A}^{(L_1)}) \ & = & \ - \,i \,\frac{\partial}{\partial \phi}
 \ - \ \frac{1}{2} \,e \,B \,(x^2 - y^2) .
\end{eqnarray}
Note that $\hat{p}^{cons}_x (\bm{A}^{(L_1)})$ takes a simpler form in accordance
with the fact that $\hat{H} (\bm{A}^{(L_1)})$ takes a separable form 
in the rectangular coordinate representation.
Then, with the choice of the $L_1$-gauge potential, one usually adopts the basis, 
in which $\hat{p}^{cons}_x (\bm{A}^{(L_1)})$ and $\hat{H} (\bm{A}^{(L_1)})$ are
simultaneously diagonalized as
\begin{eqnarray}
 \hat{p}^{cons}_x (\bm{A}^{(L_1)}) \,\vert \,\Psi^{(L_1)}_{n, k_x} \rangle \ &=& \ 
 k_x \, \vert \,\Psi^{(L_1)}_{n, k_x} \rangle ,  \\
 \hat{H} (\bm{A}^{(L_1)}) \,\vert \,\Psi^{(L_1)}_{n, k_x} \rangle \ &=& \  
 E_n \,\vert \,\Psi^{(L_1)}_{n, k_x} \rangle .
\end{eqnarray}
The explicit form of $\Psi^{(L_1)}_{n, k_x} (x,y)$ is known to be given in the form
\begin{equation}
 \Psi^{(L_1)}_{n, k_x} (x, y) \ = \ \frac{e^{\,i \,k_x \,x}}{\sqrt{2 \,\pi}} \,\,
 Y_n (y) , \label{Eq:Eigen_function_Psi_L1_n_kx}
\end{equation}
where
\begin{equation}
 Y_n (y) \ = \ N_n \,\,H_n \left( \frac{y - y_0}{l_B} \right) \,\,
 e^{\,- \,\frac{(y - y_0)^2}{2 \,l^2_B}} ,
\end{equation}
with
\begin{equation}
 N_n \ = \ \left( \frac{1}{\sqrt{\pi} \,\,2^n \,n ! \,\,l_B} \right)^{1/2},
 \ \ \ \ \ y_0 \ = \ \frac{k_x}{e \,B} \ = \ l^2_B \,k_x . 
\end{equation}
These are nothing but the familiar Landau eigen-functions in the 1st 
Landau gauge.
Although somewhat unconvensional, however, even with the choice
of the $L_1$-gauge potential $\bm{A}^{(L_1)}$, one
{\it can} choose the basis, in which $\hat{L}^{cons}_z (\bm{A}^{(L_1)})$ and
$\hat{H} (\bm{A}^{(L_1)})$ are simultaneously diagonalized as
\begin{eqnarray}
 \hat{L}^{cons}_z (\bm{A}^{(L_1)}) \,\vert \,\Psi^{(L_1)}_{n,m} \rangle \ & = & \  
 m \, \vert \,\Psi^{(L_1)}_{n,m} \rangle , \label{Eq:Psi_L1_n_m_A} 
 \label{Eq:Eigen_L_cons_z_Psi_L1_n_m} \\
 \hat{H} (\bm{A}^{(L_1)}) \,\vert \,\Psi^{(L_1)}_{n, m} \rangle \ &=& \  
 E_n \,\vert \,\Psi^{(L_1)}_{n, m} \rangle . \label{Eq:Psi_L1_n_m_B}
\end{eqnarray}
In this way, we now have totally four types of eigen-vectors of the 
Landau Hamiltonian,
\begin{equation}
 \vert \,\Psi^{(S)}_{n,m} \rangle, \ \ \ \vert \,\Psi^{(S)}_{n, k_x} \rangle, \ \ \ 
 \vert \,\Psi^{(L_1)}_{n,m} \rangle, \ \ \ \vert \,\Psi^{(L_1)}_{n, k_x} \rangle.
 \label{Eq:4_eigen_states}
\end{equation}

\noindent
A natural question is mutual relations between these four eigen-functions
of the Landau problem. To answer this question, we recall that the gauge potential 
in the symmetric gauge and the 1st Landau gauge are connected by the 
following gauge transformation :
\begin{equation}
 \bm{A}^{(L_1)} \ = \ \bm{A}^{(S)} \ + \ \nabla \chi, \ \ \ \mbox{with}
 \ \ \ \chi \ = \ - \,\frac{1}{2} \,B \,x \,y .
\end{equation}
In quantum mechanics, the above gauge transformation can be realized as a 
unitary transformation represented as
\begin{equation}
 - \,i \,\nabla \ + \ e \,\bm{A}^{(L_1)} \ = \ U \,
 \left( - \,i \,\nabla \ + \ e \,\bm{A}^{(S)} \right) \,U^\dagger,
\end{equation}
with
\begin{equation}
 U \ = \ e^{\,- \,i \,e \,\chi} \ = \ e^{\,i \,\frac{1}{2} \,e \,B \,x \,y} .
\end{equation}
It can easily be shown that
\begin{eqnarray}
 U \,\hat{H} (\bm{A}^{(S)}) \,U^\dagger \ \ &=& \ \hat{H} (\bm{A}^{(L_1)}), \\
 U \,\hat{p}^{cons}_x (\bm{A}^{(S)}) \,U^\dagger \ &=& \ 
 \hat{p}^{cons}_x (\bm{A}^{(L_1)}), \\
 U \,\hat{L}^{cons}_z (\bm{A}^{(S)}) \,U^\dagger \ &=& \ 
 \hat{L}^{cons}_z (\bm{A}^{(L_1)}).
\end{eqnarray}
Multiplying $U$ on both sides of (\ref{Eq:Eigen_L_cons_z_Psi_S_n_m}) from 
the left and inserting the identity $U^\dagger \,U = 1$, we obtain
\begin{equation}
 U \,\hat{L}^{cons}_z (\bm{A}^{(S)}) \,U^\dagger \,\,U \,
 \vert \,\Psi^{(S)}_{n,m} \rangle \ = \ m \,U \,
 \vert \,\Psi^{(S)}_{n,m} \rangle . 
\end{equation}
Then, by using  $U \,\hat{L}^{cons}_z (\bm{A}^{(S)}) \,U^\dagger = 
\hat{L}^{cons}_z (\bm{A}^{(L_1)})$, this equation becomes
\begin{equation}
 \hat{L}^{cons}_z (\bm{A}^{(L_1)}) \,U \,\vert \,\Psi^{(S)}_{n,m} \rangle
 \ = \ m \,U \,\vert \,\Psi^{(S)}_{n,m} \rangle.
\end{equation}
Comparing it with (\ref{Eq:Eigen_L_cons_z_Psi_L1_n_m}), we therefore 
conclude that (up to phase)
\begin{equation}
 \vert \,\Psi^{(L_1)}_{n,m} \rangle \ = \ U \, \vert \Psi^{(S)}_{n,m} \rangle .
 \label{Eq:UT_Psi_L1_n_m_S_n_m}
\end{equation}
Similarly, we can readily confirm the relation
\begin{equation}
 \vert \,\Psi^{(L_1)}_{n, k_x} \rangle \ = \ U \,\vert \,\Psi^{(S)}_{n, k_x} \rangle .
 \label{Eq:UT_Psi_L1_n_kx_S_n_kx}
\end{equation}
What is meant by these two relations is nothing surprising. 
First, the relation (\ref{Eq:UT_Psi_L1_n_m_S_n_m}) dictates that the two 
Landau eigen-states
$\vert \,\Psi^{(L_1)}_{n,m} \rangle$ and $\vert \,\Psi^{(S)}_{n,m} \rangle$ are related
through the gauge transformation matrix $U$. 
This indicates that these two eigen-states belong to the same family.
Similarly, the relation (\ref{Eq:UT_Psi_L1_n_kx_S_n_kx}) shows that 
$\vert \,\Psi^{(L_1)}_{n, k_x} \rangle$ and $\vert \,\Psi^{(S)}_{n, k_x} \rangle$ are
related through the gauge transformation matrix $U$, which implies that
these two eigen-states
$\vert \,\Psi^{(L_1)}_{n, k_x} \rangle$ and $\vert\,\Psi^{(S)}_{n, k_x} \rangle$ 
belong to the same family.
More generally, any eigen-states belonging to the 1st family and
any eigen-states belonging to the 2nd family can be constructed
by using the gauge-transformation martix as follows : 
\begin{equation}
 \vert \,\Psi^{(\chi)}_{n,m} \rangle \ = \ U^{(\chi)} \,
 \vert \,\Psi^{(S)}_{n,m} \rangle \ = \ 
 e^{- i \,\,e \,\chi} \,\vert \,\Psi^{(S)}_{n,m} \rangle ,
\end{equation}
and
\begin{equation}
 \vert \,\Psi^{(\chi^\prime)}_{n,k_x} \rangle \ = \ U^{(\chi^\prime)} \,
 \vert \,\Psi^{(L_1)}_{n,k_x} \rangle \ = \ 
 e^{- i \,\,e \,\chi^\prime} \,\vert \,\Psi^{(L_1)}_{n,k_x} \rangle ,
\end{equation}
with $\chi$ and $\chi^\prime$ being {\it arbitrary} harmonic functions.
What is nontrivial is the relation between the 1st class of eigen-states
$\vert \,\Psi^{(\chi)}_{n,m} \rangle$ and the 2nd class
of eigen-states $\vert \,\Psi^{(\chi^\prime)}_{n, k_x} \rangle$. 
As we shall confirm through the following analysis, it turns out that they 
are not simply related by a single gauge transformation matrix,
and that they actually belong to different family. (The meaning of the terminology
{\it family} here will become clear in the next section.)

\section{Matrix elements of 3 kinds of momentum and OAM operators
between 4 types of Landau eigen-states}
\label{sec3}

Our objective here is to evaluate the matrix elements of the 3 kinds of
momentum operators
\begin{eqnarray}
 \hat{p}^{can}_x \ &=& \ - \,i \,\frac{\partial}{\partial x}, \\
 \hat{p}^{cons}_x \ &=& \ - \,i \,\frac{\partial}{\partial x} \ + \ 
 e \,A_x \ + \ e \,B \,y , \\
 \hat{p}^{mech}_x \ &=& \ - \,i \,\frac{\partial}{\partial x} \ + \ 
 e \,A_x ,
\end{eqnarray}
and the 3 kinds of OAM operators
\begin{eqnarray}
 \hat{L}^{can}_z \ &=& \ - \,i \,\frac{\partial}{\partial \phi}, \\
 \hat{L}^{cons}_z \ &=& \ - \,i \,\frac{\partial}{\partial \phi} \ + \ 
 e \,r \,A_\phi \ - \ \frac{1}{2} \,e \,B \,r^2, \\
 \hat{L}^{mech}_z \ &=& \ - \,i \,\frac{\partial}{\partial \phi} \ + \ 
 e \,r \,A_\phi ,
\end{eqnarray}
between the 4 different types of Landau eigen-states shown by 
(\ref{Eq:4_eigen_states}).
The reason of such studies will become clear at the end of calculations.

\subsection{Matrix elements in the $\vert\,n, k_x \rangle$-basis}

We first consider the matrix elements between the eigen-states
$\vert \,\Psi^{(L_1)}_{n, k_x} \rangle$ and $\vert \,\Psi^{(S)}_{n, k_x} \rangle$,
which simultaneously diagonalize $\hat{p}^{cons}_x$ and the
Landau Hamiltonian $\hat{H}$. Hereafter, we call them the
$\vert \,n, k_x \rangle$-basis.
Remember that these two eigen-functions are related by the
gauge transformation matrix $U^\dagger$ as
$\vert \,\Psi^{(S)}_{n, k_x} \rangle = U^\dagger \,\vert \,\Psi^{(L_1)}_{n, k_x} \rangle$.

The calculations of these matrix elements are straightforward but very tedious.
We therefore describe the detailed derivation in Appedices A and B, and
show here only the final answers.
Summarized in Table 1 are the matrix elements of the three momentum 
operators as well as those of the three OAM operators in the
$\vert \,n, k_x \rangle$-basis.

\vspace{2mm}
\begin{table}[thb]
\caption{Matrix elements of the 3 kinds of momentum and OAM 
operators in the $\vert \,n , k_x \rangle$-basis.}
\renewcommand{\arraystretch}{1.2}
\begin{center}
\begin{tabular}{ccc}
\hline\hline
 $\hat{O}$ \ & \ \hspace{-26mm} 
 $\langle \Psi^{(L_1)}_{n,k_x^\prime} \,\vert \, \hat{O} (\bm{A}^{(L_1)}) \,
 \vert \, \Psi^{(L_1)}_{n,k_x} \rangle$ \ \  
 & \ \hspace{-20mm}  
 $\langle \Psi^{(S)}_{n,k_x^\prime} \,\vert \, \hat{O} (\bm{A}^{(S)}) \,\vert \, 
 \Psi^{(S)}_{n,k_x} \rangle$ \ \ \ \\
 \hline\hline
 $\hat{p}^{can}_x$ \ & \ \hspace{-26mm} $k_x \,\delta (k^\prime_x - k_x)$ \ 
 & \ \hspace{-22mm} $\frac{1}{2} \,k_x \,\delta (k^\prime_x - k_x)$ \ \\
 \hline
 $\hat{p}^{mech}_x (\bm{A})$ \ & \ \hspace{-26mm} 0 \ & \ \hspace{-22mm} 0 \ \\
 \hline
 $\hat{p}^{cons}_x (\bm{A})$ \ & \ \hspace{-26mm} $k_x \,\delta (k^\prime_x - k_x)$ \ 
 & \ \hspace{-22mm} $k_x \,\delta (k^\prime_x - k_x)$ \ \\
 \hline\hline \hspace{8mm}
 \multirow{2}{15mm}{$\hat{L}^{can}_z$} \ & \ \hspace{-2mm}
 \multirow{2}{70mm}{$\left\{\, n \, + \, \frac{1}{2} \,- \,\frac{k_x^2}{e \,B} \right\} \,\delta (k_x^\prime - k_x)$} 
 & \hspace{-16mm} 
 $\left\{\,n \,+ \,\frac{1}{2} \, - \,\frac{k_x^2}{2 \,e \,B} \right\} \,\delta (k_x^\prime - k_x)$ 
 \ \\
 & \ & \ \hspace{-18mm} $+ \,\frac{e \,B}{2} \,\delta^{\prime\prime} (k^\prime_x - k_x)$ \ \\
 \hline
 $\hat{L}^{mech}_z (\bm{A})$ \ & \ \hspace{-26mm} $(2 \,n + 1) \,\delta (k_x^\prime - k_x)$ \ & 
 \ \hspace{-24mm} $(2 \,n + 1) \,\delta (k_x^\prime - k_x)$ \ \\
 \hline
 \multirow{2}{10mm}{$\hat{L}^{cons}_z (\bm{A})$} \ \ \ \ & \hspace{-22mm} 
 $\left\{\, n \, + \, \frac{1}{2} \,- \,\frac{k_x^2}{2 \,e \,B} \right\} \,\delta (k_x^\prime - k_x)$ 
 & \hspace{-18mm}
 $\left\{\, n \, + \, \frac{1}{2} \,- \,\frac{k_x^2}{2 \,e \,B} \right\} \,\delta (k_x^\prime - k_x)$ \ \\
 \ & \ \hspace{-22mm} $+ \,\frac{e \,B}{2} \,\delta^{\prime\prime} (k^\prime_x - k_x)$ & \ 
 \hspace{-18mm} $+ \,\frac{e \,B}{2} \,\delta^{\prime\prime} (k^\prime_x - k_x)$ \ \\
\hline\hline
\end{tabular}
\end{center}
\end{table}

This table reveals several important facts. First, we notice that
\begin{eqnarray}
 \langle \Psi^{(L_1)}_{n, k^\prime_x} \,\vert \, \hat{p}^{can}_x \,
 \vert \, \Psi^{(L_1)}_{n, k_x} \rangle \ &\neq& \ 
 \langle \Psi^{(S)}_{n, k^\prime_x} \,\vert \,\hat{p}^{can}_x \,
 \vert \,\Psi^{(S)}_{n, k_x} \rangle, \\
  \langle \Psi^{(L_1)}_{n, k^\prime_x} \,\vert \, \hat{L}^{can}_z \,
 \vert \, \Psi^{(L_1)}_{n, k_x} \rangle \ &\neq& \ 
 \langle \Psi^{(S)}_{n, k^\prime_x} \,\vert \,\hat{L}^{can}_z \,
 \vert \,\Psi^{(S)}_{n, k_x} \rangle .
\end{eqnarray}
These results are nothing surprising, since the canonical quantities, 
$\hat{p}_x$ and $\hat{L}^{can}_z$, are 
widely recognized to be {\it gauge-variant} operators.
On the other hand, we see that
\begin{eqnarray}
 \langle \Psi^{(L_1)}_{n, k^\prime_x} \,\vert \, \hat{p}^{cons}_x (\bm{A}^{(L_1)}) \,
 \vert \, \Psi^{(L_1)}_{n, k_x} \rangle \ &=& \ 
 \langle \Psi^{(S)}_{n, k^\prime_x} \,\vert \,\hat{p}^{cons}_x (\bm{A}^{(S)}) \,
 \vert \,\Psi^{(S)}_{n, k_x} \rangle, \\
 \langle \Psi^{(L_1)}_{n, k^\prime_x} \,\vert \, \hat{p}^{mech}_x (\bm{A}^{(L_1)}) \,
 \vert \, \Psi^{(L_1)}_{n, k_x} \rangle \ &=& \ 
 \langle \Psi^{(S)}_{n, k^\prime_x} \,\vert \,\hat{p}^{mech}_x (\bm{A}^{(S)}) \,
 \vert \,\Psi^{(S)}_{n, k_x} \rangle, 
\end{eqnarray} 
and 
\begin{eqnarray}
 \langle \Psi^{(L_1)}_{n, k^\prime_x} \,\vert \, \hat{L}^{cons}_z (\bm{A}^{(L_1)}) \,
 \vert \, \Psi^{(L_1)}_{n, k_x} \rangle \ &=& \ 
 \langle \Psi^{(S)}_{n, k^\prime_x} \,\vert \,\hat{L}^{cons}_z (\bm{A}^{(S)}) \,
 \vert \,\Psi^{(S)}_{n, k_x} \rangle, \\
 \langle \Psi^{(L_1)}_{n, k^\prime_x} \,\vert \, \hat{L}^{mech}_z (\bm{A}^{(L_1)}) \,
 \vert \, \Psi^{(L_1)}_{n, k_x} \rangle \ &=& \ 
 \langle \Psi^{(S)}_{n, k^\prime_x} \,\vert \,\hat{L}^{mech}_z (\bm{A}^{(S)}) \,
 \vert \,\Psi^{(S)}_{n, k_x} \rangle .
\end{eqnarray}
These are also expected relations, because all these four operators
$\hat{p}^{cons}_x (\bm{A}) $, $\hat{p}^{mech}_x (\bm{A})$ and 
$\hat{L}^{cons}_z (\bm{A})$, $\hat{L}^{mech}_z (\bm{A})$
transform {\it covariantly} under a gauge transformation, and because
the two bases $\vert \,\Psi^{(S)}_{n, k_x} \rangle$ and 
$\vert \,\Psi^{(L_1)}_{n, k_x} \rangle$
belong to the {\it same} family as related by the gauge transformation matrix
$U^\dagger$ as $\vert \,\Psi^{(S)}_{n, k_x} \rangle = 
U^\dagger \,\vert \,\Psi^{(L_1)}_{n, k_x} \rangle$.

Not so obvious is the following observation. Namely,
although the mathematical properties of the conserved momentum operator 
and the mechanical momentum operator, i.e. their covariant nature under 
a gauge transformation, are entirely the same, 
Table I apparently reveals the existence of some
critical differences between those. Note that the matrix element of
the mechanical momentum operator $p^{mech}_x (\bm{A})$ is zero
in both of the eigen-states $\vert \,\Psi^{(L_1)}_{n, k_x} \rangle$ and
$\vert \,\Psi^{(S)}_{n, k_x} \rangle$. This appears to indicate {\it physical}
nature of the mechanical momentum. In fact, in classical mechanics, the 
Landau electron makes a cyclotron (or circular) motion around some center.
The position of the center of this cyclotron motion is arbitrary because of
the special nature of the Landau problem, in which the magnetic field is
uniformly spreading over the whole 2-dimensional plane. Nevertheless, the
center of this cyclotron motion is a constant in time in both of classical
mechanics and quantum mechanics, so that the time-average
or the expectation value of the physical electron momentum along the
$x$-direction must evidently be zero. 

In contrast, the physical meaning of
the conserved momentum is not necessarily obvious. From Table I, we see that
 \begin{equation}
 \langle \Psi^{(L_1)}_{n, k^\prime_x} \,\vert \,\hat{p}^{can}_x \,\vert \,\Psi^{(L_1)}_{n, k_x} \rangle
 \ = \ \langle \Psi^{(L_1)}_{n, k^\prime_x} \,\vert \,\hat{p}^{cons}_x (\bm{A}^{(L_1)}) \,\vert
 \,\Psi^{(L_1)}_{n, k_x} \rangle ,
\end{equation}
but
\begin{equation}
 \langle \Psi^{(S)}_{n, k^\prime_x} \,\vert \,\hat{p}^{can}_x \,\vert \,\Psi^{(S)}_{n, k_x} \rangle
 \ \neq \ \langle \Psi^{(S)}_{n, k^\prime_x} \,\vert \,\hat{p}^{cons}_x (\bm{A}^{(S)}) \,\vert
 \,\Psi^{(S)}_{n, k_x} \rangle .
 \end{equation}
The first equality is understandable because $\hat{p}^{cons}_x$ just
reduces to $\hat{p}^{can}_x$ with the choice of the $L_1$-gauge potential.
The second inequality however shows that, in general, the matrix elements
of $\hat{p}^{cons}_x$ and $\hat{p}^{can}_x$ do not have any simple relation.
One may also notice that the matrix elements of  $\hat{p}^{cons}_x$ and 
$\hat{p}^{can}_x$ in the $\vert \,n, k_x \rangle$-basis are proportional to the quantum 
number $k_x$.
In the framework of quantum mechanics, $k_x$ is an important quantum
number, which characterizes the Landau eigen-states in the 
$\vert \,n, k_x \rangle$-basis.
Nonetheless, one must recognize the fact that this quantum number is not such
a quantity, which has a direct connection with observables of the Landau electron. 
(This is also indicated by the fact that this quantum number never appears 
in the eigen-functions in the symmetric gauge.)

Also puzzling is fairly clumsy expression for the matrix element
of the conserved OAM operator $\hat{L}^{cons}_z (\bm{A})$,
which should be contrasted with the very simple form for the matrix element 
of the mechanical OAM operator $\hat{L}^{mech}_z (\bm{A})$.
In fact, the following equalities 
\begin{eqnarray}
 \langle \Psi^{(L_1)}_{n, k^\prime_x} \,\vert \,\hat{L}^{mech}_z (\bm{A}^{(L_1)}) \,
 \vert \,\Psi^{(L_1)}_{n, k_x} \rangle \ &=& \ 
 \langle \Psi^{(S)}_{n, k^\prime_x} \,\vert \,\hat{L}^{mech}_z (\bm{A}^{(S)}) \,
 \vert \,\Psi^{(S)}_{n, k_x} \rangle \nonumber \\
 \ &=& \ (2 \,n \ + \ 1) \,\delta (k^\prime_x - k_x) , \label{Eq:mat_kx}
\end{eqnarray}
for the matrix elements of the mechanical OAM is already indicating close 
resemblance with the familiar answer for the expectation value of the same operator
between the Landau eigen-states $\vert \,\Psi^{(S)}_{n,m} \rangle$ in the symmetric
gauge given as
\begin{equation}
  \langle \Psi^{(S)}_{n, m} \,\vert \,\hat{L}^{mech}_z (\bm{A}^{(S)}) \,
  \vert \,\Psi^{(S)}_{n,m} \rangle \ = \ 2 \,n \ + \ 1 .
\end{equation}
The difference is the appearance of the Dirac's delta function
$\delta (k^\prime_x - k_x)$ in the matrix elements given by (\ref{Eq:mat_kx}), 
which comes from the non-normalizable nature of the $\vert \,n, k_x \rangle$
basis functions. However, corresponding normalizable basis functions can easily
be constructed, if we consider wave-packet states.
To confirm it, we first recall that the Landau eigen-functions in the $L_1$-gauge
can be expressed as a product of non-normarizable plane-wave function
$f_{k_x} (x)$ and normalizable functions $Y_n (y)$ as
\begin{equation}
 \Psi^{(L_1)}_{n, k_x} (x,y) \ = \ f_{k_x} (x) \,Y_n (y) ,
\end{equation}
with $f_{k_x} (x) = (1 / \sqrt{2 \,\pi})\,e^{\,i \,k_x \,x}$.
We now replace the above plane-wave functions by wave-packets as
\begin{eqnarray}
 f_{k_x} (x) \ \rightarrow \ F_{k_x} (x) &\equiv& 
 \int_{- \,\infty}^\infty \,g (k - k_x) \,f_k (x) \,d k \nonumber \\ 
 &=& 
 \int_{- \,\infty}^\infty \,\frac{dk}{\sqrt{2 \,\pi}} \,\,g(k - k_x) \,e^{\,i \,k \,x} .
\end{eqnarray}
Here, $g (k)$ is an appropriate weight function of superposition, which has a 
peak at $k = 0$, and is normalized as, 
\begin{equation}
 \int_{-\,\infty}^\infty \,\,d k \,\,\left[ g (k) \right]^2 \ = \ 1.
\end{equation}
In correspondence with the above replacement, the eigen-functions in the
$L_1$-gauge are also replaced as follows : 
\begin{equation}
 \Psi^{(L_1)}_{n, k_x} (x,y) \ \rightarrow \ \tilde{\Psi}^{(L_1)}_{n, k_x} (x,y)
 \ \equiv \ F_{k_x} (x) \,Y_n (y).
\end{equation}
On account of the definition of $F_{k_x} (x)$, the normalizable functions 
above can also be expressed as
\begin{equation}
 \tilde{\Psi}^{(L_1)}_{n, k_x} (x,y) \ = \ \int_{- \,\infty}^\infty \,
 dk \,\,g (k - k_x) \,\Psi^{(L_1)}_{n, k} (x,y) .
\end{equation}
Now, it is an easy exercise to show the equality
\begin{equation}
 \langle \tilde{\Psi}^{(L_1)}_{n, k_x} \,\vert\,L^{mech}_z (\bm{A}^{(L_1)}) \,
 \vert \,\tilde{\Psi}^{(L_1)}_{n, k_x} \rangle \ = \ 2 \,n \ + \ 1.
\end{equation}
It is a little surprising that this simple relation has never been written down before.
It must be a highly nontrivial finding, since it shows that
the expectation value of the mechanical OAM
operator precisely coincides between both of the symmetric gauge eigen-states
and the 1st Landau gauge eigen-states.
 (It is also obvious that this property can be extended to any eigen-states
with {\it arbitrary} gauge potentials.) 
Surprisingly and importantly, this
gauge-class-independence of the matrix elements does not hold for the
conserved OAM operator, even though the mechanical OAM
operator and the conserved OAM operator have exactly the same
covariant gauge transformation property under an arbitrary gauge
transformation. (See the next section for further discussion.)

To sum up, the consideration above already indicates the existence of critical 
physical difference between the conserved quantities and the mechanical
ones, despite the fact that both transform {\it covariantly} under a gauge transformation. 
We shall pursue this nontrivial observation further through the investigation 
of the matrix elements of the same operators in the $\vert \,n, m \rangle$-basis 
in the next subsection.  

\subsection{Matrix elements in the $\vert \,n, m \rangle$-basis}

In this subsection, we evaluate the matrix elements of the three momentum
operators and the three OAM operators between the two Landau eigen-states
$\vert \,\Psi^{(S)}_{n,m} \rangle$ and $\vert \,\Psi^{(L_1)}_{n,m} \rangle$ belonging to the
$\vert \,n,m \rangle$-basis class.
Since these two eigen-states are related as $\vert \,\Psi^{(L_1)}_{n,m} \rangle = 
U \,\vert \,\Psi^{(S)}_{n,m} \rangle$, our main task below is to evaluate the matrix
elements of the above operators between the symmetric-gauge
eigen-states $\vert \,\Psi^{(S)}_{n,m} \rangle$.
Since the diagonal matrix elements $\langle \Psi^{(S)}_{n,m} \,\vert \,O (\bm{A}^{(S)}) \,\vert \,
\Psi^{(S)}_{n,m} \rangle$ were already investigated in many previous papers, we
extend these analyses to the non-diagonal matrix elements
$\langle \Psi^{(S)}_{n,m^\prime} \,\vert \,O (\bm{A}^{(S)}) \,\vert \,\Psi^{(S)}_{n,m} \rangle$ 
with respect to the magnetic quantum number $m$. 
This is partially motivated by the fact that
we had to deal with the non-diagonal matrix elements
$\langle \Psi^{(L_1)}_{n, k^\prime_x} \,\vert \,O (\bm{A}^{(L_1)}) \,\vert \,\Psi^{(L_1)}_{n, k_x} \rangle$
in the corresponding calculations in the $\vert \,n, k_x \rangle$-basis, because of the nature of
the delta-function normalization of the eigen-states $\vert \,\Psi^{(L_1)}_{n, k_x} \rangle$
along the $x$-direction.
To evaluate the non-diagonal matrix elements
$\langle \Psi^{(S)}_{n,m^\prime} \,\vert \,O (\bm{A}^{(S)}) \,\vert \,\Psi^{(S)}_{n,m} \rangle$,
we find it much easier to use the algebraic method rather than analytical method.
In the following, we therefore prepare necessary algebraic formulation
to handle the problem.

First, besides the familiar mechanical momentum
$\bm{\Pi} = \bm{p} \,+ \,e \,\bm{A} (\bm{r})$, it is convenient to introduce the 
following quantity, which was called in \cite{Goerbig2009} the pseudo-momentum : 
\begin{equation}
 \tilde{\bm{\Pi}} \ \equiv \ \bm{p} \ - \ e \,\bm{A} (\bm{r}) .
\end{equation}
(Note that the pseudo-momentum above has nothing to do with the
quantity with the same name appearing in \cite{Yoshioka2002}
\nocite{Konstantinou2016}-\cite{Konstantinou2017}.)
Clearly, this quantity does not transform covariantly under
a gauge transformation, and, as we shall see shortly,  it is useful only 
in the symmetric gauge. Here and hereafter,
we omit the {\it hat} symbol for the quantum operator including
the momentum operator $\hat{\bm{p}}$ to avoid notational complexity. 
Note that, with use of the quantities $\bm{\Pi}$ and $\tilde{\bm{\Pi}}$, 
the canonical momentum and the vector potential can be expressed 
as \cite{Goerbig2009}
\begin{equation}
 \bm{p} \ = \ \frac{1}{2} \,\left( \bm{\Pi} \ + \ \tilde{\bm{\Pi}} \right), \ \ \ \ \ 
 \bm{A} (\bm{r}) \ = \ \frac{1}{2 \,e} \,
 \left( \bm{\Pi} \ - \ \tilde{\bm{\Pi}} \right) .
 \end{equation}
It is easy to verify the following commutation relations (C.R.s) : 
\begin{equation}
 \left[ \Pi_x, \Pi_y \right] \ = \ - \,i \,\frac{1}{l^2_B} , \ \ \ \ \ 
 \left[ \tilde{\Pi}_x, \tilde{\Pi}_y \right] \ = \ + \,i \,\frac{1}{l^2_B} .
\end{equation}
We can also show that
\begin{eqnarray}
 \left[ \Pi_x, \tilde{\Pi}_x \right] \ &=& \ 2 \,i \,e \,\frac{\partial A_x}{\partial x}, \\
 \left[ \Pi_y, \tilde{\Pi}_y \right] \ &=& \ 2 \,i \,e \,\frac{\partial A_y}{\partial y}, \\
 \left[ \Pi_x, \tilde{\Pi}_y \right] \ &=& \ - \,\left[ \tilde{\Pi}_x, \Pi_y \right]
 \ = \ i \,e \,\left( \frac{\partial A_x}{\partial y} \ + \ 
 \frac{\partial A_y}{\partial x} \right) .
\end{eqnarray}
These mixed C.R.s also mean that $\tilde{\Pi}_x$ and $\tilde{\Pi}_y$ do not commute
with the Landau Hamiltonian :
\begin{equation}
 \left[ \tilde{\Pi}_x, H \right] \ \neq \ 0, \ \ \ \ 
 \left[ \tilde{\Pi}_y, H \right] \ \neq \ 0 .
\end{equation}
However, the above unwanted mixed commutators can be avoided with a particular
choice of gauge, i.e. with the choice of the symmetric gauge potential,
$\bm{A} (\bm{r}) \rightarrow \bm{A}^{(S)} (\bm{r}) = \frac{1}{2} \,( -\,y, x)$.
In fact, in this special choice of gauge potential, it holds that
\begin{equation}
 \left[ \Pi_x, \tilde{\Pi}_x \right] \ = \ \left[ \Pi_y, \tilde{\Pi}_y \right] \ = \ 
 \left[ \Pi_x, \tilde{\Pi}_y \right] \ = \ \left[ \tilde{\Pi}_x, \Pi_y \right] \ = \ 0 ,
\end{equation}
which also means that
$\left[ \tilde{\Pi}_x, H \right] \, = \, \left[ \tilde{\Pi}_y, H \right] \, = \, 0$.

Now, it is convenient to introduce two kinds of ladder operator by
\begin{eqnarray}
 a \ &=& \ i \,\frac{l_B}{\sqrt{2}} \,\left( \Pi_x \ - \ i \,\Pi_y \right), \ \ \ \ \ \,\, 
 a^\dagger \ = \ - \,i \,\frac{l_B}{\sqrt{2}} \,\left( \Pi_x \ + \ i \,\Pi_y \right), \\
 b \ &=& \ i \,\frac{l_B}{\sqrt{2}} \,
 \left( \tilde{\Pi}_x \ + \ i \,\tilde{\Pi}_y \right), \ \ \ \ \ 
 b^\dagger \ = \ - \,i \,\frac{l_B}{\sqrt{2}} \,
 \left( \tilde{\Pi}_x \ - \ i \,\tilde{\Pi}_y \right) .
\end{eqnarray}
They satisfy the following C.R.'s : 
\begin{eqnarray}
 &\,& \left[ a, a^\dagger \right] \ = \ 1, \ \ \ \left[ b, b^\dagger \right] \ = \ 1, \\
 &\,& \left[ a, b \right] \ = \ \left[ a, b^\dagger \right] \ = \ 
 \left[ a^\dagger, b \right] \ = \ \left[ a, b^\dagger \right] \ = \ 0 , 
\end{eqnarray}
Since we have
\begin{eqnarray}
 \Pi_x \ &=& \ p_x \ - \ \frac{1}{2} \,e \,B \,y, \ \ \ \ \ 
 \Pi_y \ = \ p_y \ + \ \frac{1}{2} \,e \,B \,x , \\
 \tilde{\Pi}_x \ &=& \ p_x \ + \ \frac{1}{2} \,e \,B \,y, \ \ \ \ \ 
 \tilde{\Pi}_y \ = \ p_y \ - \ \frac{1}{2} \,e \,B \,x ,
\end{eqnarray}
in the symmetric gauge, we can express as
\begin{eqnarray}
 p_x \ &=& \ \frac{1}{2} \,\left( \Pi_x \ + \ \tilde{\Pi}_x \right), \ \ \ \ \ 
 p_y \ = \ \frac{1}{2} \,\left( \Pi_y \ + \ \tilde{\Pi}_y \right), \\
 x \ &=& \ l^2_B \, \left( \Pi_y \ - \ \tilde{\Pi}_y \right), \ \ \ \ \
 y \ = \ - \,l^2_B \, \left( \Pi_x \ - \ \tilde{\Pi}_x \right) . 
\end{eqnarray}
Thus, $\bm{\Pi}$ and $\tilde{\bm{\Pi}}$ can eventually be expressed with the
ladder operators as
\begin{eqnarray}
 \Pi_x \ &=& \ - \,i \,\frac{1}{\sqrt{2} \,l_B} \,\left( a \ - \ a^\dagger \right), \ \ \ \ \, 
 \Pi_y \ = \ + \,\frac{1}{\sqrt{2} \,l_B} \,\left( a \ + \ a^\dagger \right), \\
 \tilde{\Pi}_x \ &=& \ - \,i \,\frac{1}{\sqrt{2} \,l_B} \,\left( b \ - \ b^\dagger \right), \ \ \ \ \ 
 \tilde{\Pi}_y \ = \ - \,\frac{1}{\sqrt{2} \,l_B} \,\left( b \ + \ b^\dagger \right) .
\end{eqnarray}
The three momenta of our interest are then expressed with the ladder operators as
\begin{eqnarray}
 \hspace{-10mm}
 p^{can}_x \ \ \ \ &=& \ \frac{1}{2} \,\left( \Pi_x \ + \ \tilde{\Pi}_x \right) \ = \ 
 - \,i \,\frac{1}{2 \,\sqrt{2} \,l_B} \,\left( a \ - \ a^\dagger \ + \ b \ - \ b^\dagger \right), 
 \hspace{2mm} \\
 \hspace{-10mm}
 p^{cons}_x (\bm{A}^{(S)}) &=& \ \tilde{\Pi}_x \ = \ - \,i \,\frac{1}{\sqrt{2} \,l_B} \,
 \left( b \ - \ b^\dagger \right) , \label{Eq:p_cons_x_ladder} \\
 \hspace{-10mm}
 p^{mech}_x (\bm{A}^{(S)}) &=& \ \Pi_x \ = \ - \,i \,\frac{1}{\sqrt{2} \,l_B} \,
 \left( a \ - \ a^\dagger \right) . \label{Eq:p_mech_x_ladder}
\end{eqnarray}
Similarly, the three OAMs of our interest can be expressed as
\begin{eqnarray}
 L^{can}_z \ \ \ \ \ &=& \ x \,p_y \ - \ y \,p_x \ = \ 
 a^\dagger \,a \ - \ b^\dagger \,b , \label{Eq:L_can_z_ladder} \\
 L^{cons}_z (\bm{A}^{(S)}) \ &=& \ \ \ L^{can}_z \  = \  
 a^\dagger \,a \ - \ b^\dagger \,b , \label{Eq:L_cons_z_ladder} \\
 L^{mech}_z (\bm{A}^{(S)}) \ &=& \ x \,\Pi_y \ - \ y \,\Pi_x \ = \ 
 2 \,a^\dagger \,a \ + \ 1 \ + \ ( a \,b \ + \ a^\dagger \,b^\dagger ) .
 \hspace{6mm}
 \label{Eq:L_mech_z_ladder}
\end{eqnarray}
Finally, the Landau eigen-state $\vert \,\Psi^{(S)}_{n,m} \rangle$
in the symmetric gauge are represented as
\begin{equation}
 \vert \,\Psi^{(S)}_{n,m} \rangle \ = \ \vert \,n \rangle^A \,\,\vert \,n - m \rangle^B
\end{equation}
with $n$ and $m$ being integers satisfying the constraint $m \leq n$. 
Here, $\vert \,n \rangle^A$ and $\vert \,n^\prime \rangle^B$
are the eigen-states of the Harmonic oscillator, respectively corresponding to
the creation operators $a^\dagger$ and $b^\dagger$, represented as
\begin{equation}
 \vert \,n \rangle^A \ = \ \frac{(a^\dagger)^n}{\sqrt{n \,!}} \,\vert \,0 \rangle , \ \ \ 
 \vert \,n^\prime \rangle^B \ = \ \frac{(b^\dagger)^{n^\prime}}{\sqrt{n^\prime \,!}} \,\vert \,0 \rangle .
\end{equation}
In the following, we can also use the following familiar identities : 
\begin{eqnarray}
 a^\dagger \,\vert \,n \rangle^A \ &=& \ \sqrt{n + 1} \,\vert \, n + 1 \rangle^A, \ \ \ \,\,\,
 a \,\vert \,n \rangle^A \ = \ \sqrt{n} \,\vert \, n - 1 \rangle^A, \\
 b^\dagger \,\vert \,n^\prime \rangle^B \ &=& \ 
 \sqrt{n^\prime + 1} \,\vert \, n^\prime  + 1 \rangle^B, \ \ \ 
 b \,\vert \,n^\prime \rangle^B \ = \ \sqrt{n^\prime} \,\vert \, n^\prime - 1 \rangle^B.
\end{eqnarray}
We are now ready to evaluate the required matrix elements in the $\vert \,n,m \rangle$-basis.
Although the calculations are straightforward, we describe them in Appendix C for
the sake of completeness.
Summarized in Table 2 are the matrix elements of the three momentum operators
and the three OAM operators in the $\vert \,n, m \rangle$-basis.
Table 1 and Table 2 together with their interpretations given below are the 
central achievement of the present paper.

\vspace{2mm}
\begin{table}[thb]
\caption{Matrix elements of three momenta and OAMs in $\vert \,n, m \rangle$-basis}
\renewcommand{\arraystretch}{1.2}
\begin{center}
\begin{tabular}{ccc}
\hline\hline
 \ $\hat{O}$ \ & \ \ \hspace{-2mm}   
 $\langle \Psi^{(L_1)}_{n,m^\prime} \,\vert \, \hat{O} (\bm{A}^{(L_1)}) \,\vert \, \Psi^{(L_1)}_{n,m} \rangle$ \ \ \  
 & \hspace{-2mm} 
 $\langle \Psi^{(S)}_{n,m^\prime} \,\vert \, \hat{O} (\bm{A}^{(S)}) \,\vert \, \Psi^{(S)}_{n,m} \rangle$ \ \ \ \\
 \hline\hline
 \ \multirow{2}{10mm}{$\hat{p}^{can}_x$} \ & \ \hspace{-12mm}
 $- \,i \,\sqrt{\frac{e \,B}{2}} \,
 \Bigl\{ \sqrt{n - m} \,\delta_{m^\prime, m+1}$ \ & \ \hspace{-10mm}
 $- \,i \,\frac{1}{2} \,\sqrt{\frac{e \,B}{2}} \,
 \Bigl\{ \sqrt{n - m} \,\,\delta_{m^\prime, m+1}$ \ \\
 \ & \ \hspace{0mm} $- \,\sqrt{n-m+1} \,\,\delta_{m^\prime, m-1} \Bigr\}$ \ & \ 
 \hspace{2mm} $- \,\sqrt{n-m+1} \,\,\delta_{m^\prime, m-1} \Bigr\}$ \ \\
 \hline
 \ $\hat{p}^{mech}_x (\bm{A})$ \ & 0 & \hspace{-20mm} 0 \ \\
 \hline
 \ \multirow{2}{10mm}{$\hat{p}^{cons}_x (\bm{A})$} \ \ \ & \ \hspace{-13mm}
 $- \,i \,\sqrt{\frac{e \,B}{2}} \,
 \Bigl\{ \sqrt{n - m} \,\,\delta_{m^\prime, m+1}$ \ & \ \hspace{-14mm}
 $- \,i \,\sqrt{\frac{e \,B}{2}} \,
 \Bigl\{ \sqrt{n - m} \,\,\delta_{m^\prime, m+1}$ \ \\
 \ & \ \hspace{0mm} $- \,\sqrt{n-m+1} \,\,\delta_{m^\prime, m-1} \Bigr\}$ \ & \ 
 \hspace{-2mm} $- \,\sqrt{n-m+1} \,\,\delta_{m^\prime, m-1} \Bigr\}$ \ \\
 \hline\hline
 \ \multirow{3}{10mm}{$\hat{L}^{can}_z$} \ & \ \hspace{-6mm}
 \ $m \,\delta_{m^\prime, m}$ \ & \ \multirow{3}{10mm}{$m \,\delta_{m^\prime, m}$} \ \\
 \ & \ \hspace{-4mm} $+ \, \frac{1}{2} \,\Bigl\{ \sqrt{(n - m) \,(n - m - 1)} \,\,\delta_{m^\prime, m+2}$ \ & \ \\
 \ & \ \hspace{-0mm} $+ \,\sqrt{(n - m + 1) \,(n - m + 2)} \,\,\delta_{m^\prime, m- 2} \Bigr\}$ \ & \ \\
 \hline
 \ $\hat{L}^{mech}_z (\bm{A})$ \ & \ $(2 \,n + 1)\,\delta_{m^\prime, m}$ \ & 
 \ $(2 \,n + 1) \,\delta_{m^\prime, m}$ \ \\
 \hline
 \ $\hat{L}^{cons}_z (\bm{A})$ \ & \ $m \,\,\delta_{m^\prime, m}$ \ & \ 
 $m \,\,\delta_{m^\prime, m}$ \ \\
 \hline\hline
\end{tabular}
\end{center}
\end{table}

First, from Table 2, we confirm the following relations.
\begin{eqnarray}
 \langle \Psi^{(L_1)}_{n, m^\prime} \,\vert \, \hat{p}^{can}_x \,
 \vert \, \Psi^{(L_1)}_{n, m} \rangle \ &\neq& \ 
 \langle \Psi^{(S)}_{n, m\prime} \,\vert \,\hat{p}^{can}_x \,
 \vert \,\Psi^{(S)}_{n, m} \rangle, \\
 \langle \Psi^{(L_1)}_{n, m^\prime} \,\vert \, \hat{L}^{can}_z \,
 \vert \, \Psi^{(L_1)}_{n, k_x} \rangle \ &\neq& \ 
 \langle \Psi^{(S)}_{n, m^\prime} \,\vert \,\hat{L}^{can}_z \,
 \vert \,\Psi^{(S)}_{n, m} \rangle ,
\end{eqnarray}
while
\begin{eqnarray}
 \langle \Psi^{(L_1)}_{n, m^\prime} \,\vert \, \hat{p}^{cons}_x (\bm{A}^{(L_1)}) \,
 \vert \, \Psi^{(L_1)}_{n,m} \rangle \ &=& \ 
 \langle \Psi^{(S)}_{n, m^\prime} \,\vert \,\hat{p}^{cons}_x (\bm{A}^{(S)}) \,
 \vert \,\Psi^{(S)}_{n, m} \rangle, \\
 \langle \Psi^{(L_1)}_{n, m^\prime} \,\vert \, \hat{p}^{mech}_x (\bm{A}^{(L_1)}) \,
 \vert \, \Psi^{(L_1)}_{n, m} \rangle \ &=& \ 
 \langle \Psi^{(S)}_{n, m^\prime} \,\vert \,\hat{p}^{mech}_x (\bm{A}^{(S)}) \,
 \vert \,\Psi^{(S)}_{n, m} \rangle, 
\end{eqnarray} 
and 
\begin{eqnarray}
 \langle \Psi^{(L_1)}_{n, m^\prime} \,\vert \, \hat{L}^{cons}_z (\bm{A}^{(L_1)}) \,
 \vert \, \Psi^{(L_1)}_{n, m} \rangle \ &=& \ 
 \langle \Psi^{(S)}_{n, m^\prime} \,\vert \,\hat{L}^{cons}_z (\bm{A}^{(S)}) \,
 \vert \,\Psi^{(S)}_{n, m} \rangle, \\
 \langle \Psi^{(L_1)}_{n, m^\prime} \,\vert \, \hat{L}^{mech}_z (\bm{A}^{(L_1)}) \,
 \vert \, \Psi^{(L_1)}_{n, m} \rangle \ &=& \ 
 \langle \Psi^{(S)}_{n, m^\prime} \,\vert \,\hat{L}^{mech}_z (\bm{A}^{(S)}) \,
 \vert \,\Psi^{(S)}_{n, m} \rangle .
\end{eqnarray}
The reasonable nature of these inequalities and equalities 
was already explained for the similar relations
observed for the matrix elements in the $\vert n, k_x \rangle$-basis.
(See the explanation given at the end of the previous subsection.)

Truly new insight can be obtained by comparing
Table 1 and Table 2. Let us first compare the matrix elements of the
momentum operators. We find that, for the matrix elements of the
canonical and conserved momentum operator in the $\vert \,n, k_x \rangle$-basis
and the $\vert \,n, m \rangle$-basis, there is no simple
correspondence between them :
\begin{eqnarray}
 \langle \Psi^{(L_1)}_{n, k^\prime_x} \,\vert \, \hat{p}^{can}_x \,
 \vert \, \Psi^{(L_1)}_{n, k_x} \rangle \ \ &\nLeftrightarrow& \ \ 
 \langle \Psi^{(L_1)}_{n, m^\prime} \,\vert \,\hat{p}^{can}_x \,
 \vert \,\Psi^{(L_1)}_{n, m} \rangle, \\
 \langle \Psi^{(L_1)}_{n, k^\prime_x} \,\vert \, \hat{p}^{cons}_x (\bm{A}^{(L_1)}) \,
 \vert \, \Psi^{(L_1)}_{n, k_x} \rangle \ \ &\nLeftrightarrow& \ \ 
 \langle \Psi^{(L_1)}_{n, m^\prime} \,\vert \,\hat{p}^{cons}_x (\bm{A}^{(L_1)}) \,
 \vert \,\Psi^{(L_1)}_{n, m} \rangle, \\
 \langle \Psi^{(S)}_{n, k^\prime_x} \,\vert \, \hat{p}^{can}_x \,
 \vert \, \Psi^{(S)}_{n, k_x} \rangle \ \ &\nLeftrightarrow& \ \ 
 \langle \Psi^{(S)}_{n, m^\prime} \,\vert \,\hat{p}^{can}_x \,
 \vert \,\Psi^{(S)}_{n, m} \rangle, \\
 \langle \Psi^{(S)}_{n, k^\prime_x} \,\vert \, \hat{p}^{cons}_x (\bm{A}^{(S)}) \,
 \vert \, \Psi^{(S)}_{n, k_x} \rangle \ \ &\nLeftrightarrow& \ \ 
 \langle \Psi^{(S)}_{n, m^\prime} \,\vert \,\hat{p}^{cons}_x (\bm{A}^{(S)}) \,
 \vert \,\Psi^{(S)}_{n, m} \rangle .
\end{eqnarray}
This is in some sense nothing surprising, in view of the fact that the 
$\vert \,n, k_x \rangle$-basis
and $\vert \,n, m \rangle$-basis are characterized by totally different quantum
numbers $k_x$ and $m$. However, highly nontrivial observation here is that the
matrix elements of the mechanical momentum 
operator are exactly zero in both of the $\vert n, k_x \rangle$- and 
$\vert \,n, m \rangle$-basis eigen-states :
\begin{eqnarray}
 &\,& \langle \Psi^{(L_1)}_{n, k^\prime_x} \,\vert \, \hat{p}^{mech}_x (\bm{A}^{(L_1)}) \,
 \vert \, \Psi^{(L_1)}_{n, k_x} \rangle \ = \ 
 \langle \Psi^{(L_1)}_{n, m^\prime} \,\vert \,\hat{p}^{mech}_x (\bm{A}^{(L_1)}) \,
 \vert \,\Psi^{(L_1)}_{n, m} \rangle \nonumber \\
 \hspace{-15mm}
 &=& \langle \Psi^{(S)}_{n, k^\prime_x} \,\vert \, \hat{p}^{mech}_x (\bm{A}^{(S)}) \,
 \vert \, \Psi^{(S)}_{n, k_x} \rangle \ = \ 
 \langle \Psi^{(S)}_{n, m^\prime} \,\vert \,\hat{p}^{mech}_x (\bm{A}^{(S)}) \,
 \vert \,\Psi^{(S)}_{n, m} \rangle \ = \ 0. \hspace{8mm}
\end{eqnarray}
Undoubtedly, these relations imply {\it physical nature} of the mechanical (or kinetic)
momentum as compared with the canonical momentum as well as the
conserved momentum. (The terminology {\it physical} here means the observable
nature of the quantity in question.) 
In fact, we have already explained the physical or dynamical
reason in the previous subsection why the expectation value of the mechanical 
momentum along the $x$-direction (and also the $y$-direction) vanishes exactly.  

Next, let us compare the matrix elements of the OAM operators
in the $\vert \,n, k_x \rangle$-basis and the $\vert \,n, m \rangle$-basis. 
Just like the momentum operators, we confirm that, for the matrix 
elements of the canonical and conserved OAM operators 
in the $\vert \,n, k_x \rangle$-basis
and the $\vert \,n, m \rangle$-basis, there is no physically meaningful
correspondence between them,
\begin{eqnarray}
 \langle \Psi^{(L_1)}_{n, k^\prime_x} \,\vert \, \hat{L}^{can}_z \,
 \vert \, \Psi^{(L_1)}_{n, k_x} \rangle \ \ &\nLeftrightarrow& \ \ 
 \langle \Psi^{(L_1)}_{n, m^\prime} \,\vert \,\hat{L}^{can}_z \,
 \vert \,\Psi^{(L_1)}_{n, m} \rangle, \\
 \langle \Psi^{(L_1)}_{n, k^\prime_x} \,\vert \, \hat{L}^{cons}_z (\bm{A}^{(L_1)}) \,
 \vert \, \Psi^{(L_1)}_{n, k_x} \rangle \ \ &\nLeftrightarrow& \ \ 
 \langle \Psi^{(L_1)}_{n, m^\prime} \,\vert \,\hat{L}^{cons}_z (\bm{A}^{(L_1)}) \,
 \vert \,\Psi^{(L_1)}_{n, m} \rangle, \\
 \langle \Psi^{(S)}_{n, k^\prime_x} \,\vert \, \hat{L}^{can}_z \,
 \vert \, \Psi^{(S)}_{n, k_x} \rangle \ \ &\nLeftrightarrow& \ \ 
 \langle \Psi^{(S)}_{n, m^\prime} \,\vert \,\hat{L}^{can}_z \,
 \vert \,\Psi^{(S)}_{n, m} \rangle, \\
 \langle \Psi^{(S)}_{n, k^\prime_x} \,\vert \, \hat{L}^{cons}_z (\bm{A}^{(S)}) \,
 \vert \, \Psi^{(S)}_{n, k_x} \rangle \ \ &\nLeftrightarrow& \ \ 
 \langle \Psi^{(S)}_{n, m^\prime} \,\vert \,\hat{L}^{cons}_z (\bm{A}^{(S)}) \,
 \vert \,\Psi^{(S)}_{n, m} \rangle .
\end{eqnarray}
However, there is an apparent correspondence between the matrix elements
of the mechanical OAM operator between the $\vert \,n, k_x \rangle$-basis
and the $\vert \,n , m \rangle$-basis as  
\begin{eqnarray}
 \langle \Psi^{(L_1)}_{n, k^\prime_x} \,\vert \, \hat{L}^{mech}_z (\bm{A}^{(L_1)}) \,
 \vert \, \Psi^{(L_1)}_{n, k_x} \rangle &=& 
 \langle \Psi^{(S)}_{n, k^\prime_x} \,\vert \, \hat{L}^{mech}_z (\bm{A}^{(S)}) \,
 \vert \, \Psi^{(S)}_{n, k_x} \rangle \nonumber \\
 \ &=& \ (2 \,n + 1) \,\delta (k^\prime_x - k_x), 
 \ \ \ \ \ 
\end{eqnarray}
and
\begin{eqnarray}
 \langle \Psi^{(L_1)}_{n, m^\prime} \,\vert \, \hat{L}^{mech}_z (\bm{A}^{(L_1)}) \,
 \vert \, \Psi^{(L_1)}_{n, m} \rangle \ &=& \
 \langle \Psi^{(S)}_{n, m^\prime} \,\vert \,\hat{L}^{mech}_z (\bm{A}^{(S)}) \,
 \vert \,\Psi^{(S)}_{n, m} \rangle \nonumber \\
 \ &=& \ (2 \,n + 1) \,\delta_{m^\prime,m}.
\end{eqnarray}
As already explained at the end of subsection 3.1, the main
difference comes from the non-normalizable plane-wave nature of the
basis functions $\vert \,n, k_x \rangle$.
If we introduce the normalizable wave-packet states 
$\vert \,\tilde{\Psi}^{(L_1)}_{n,k_x} \rangle$ and $\vert \,\tilde{\Psi}^{(S)}_{n,k_x} \rangle$
respectively corresponding to the states $\vert \,\Psi^{(L_1)}_{n,k_x} \rangle$ 
and $\vert \,\Psi^{(S)}_{n,k_x} \rangle$, we are led to a beautiful relationship
for the expectation values (diagonal matrix elements) of the mechanical 
OAM operator given as
\begin{eqnarray}
 \hspace{-0mm}
 &\,& \langle \tilde{\Psi}^{(L_1)}_{n, k_x} \,\vert \, \hat{L}^{mech}_z (\bm{A}^{(L_1)}) \,
 \vert \, \tilde{\Psi}^{(L_1)}_{n, k_x} \rangle \ = \ 
 \langle \tilde{\Psi}^{(S)}_{n, k_x} \,\vert \, \hat{L}^{mech}_z (\bm{A}^{(S)}) \,
 \vert \, \tilde{\Psi}^{(S)}_{n, k_x} \rangle \nonumber \\
 \hspace{-0mm}
 \ &=& \  
 \langle \Psi^{(L_1)}_{n, m} \,\vert \, \hat{L}^{mech}_z (\bm{A}^{(L_1)}) \,
 \vert \, \Psi^{(L_1)}_{n, m} \rangle \ \ = \
 \langle \Psi^{(S)}_{n, m} \,\vert \,\hat{L}^{mech}_z (\bm{A}^{(S)}) \,
 \vert \,\Psi^{(S)}_{n, m} \rangle \hspace{6mm} \nonumber \\
 \ &=& \hspace{12mm} (2 \,n + 1) .  
\end{eqnarray}
In our opinion, this is interpreted as showing {\it genuinely} gauge-invariant 
nature of the expectation value of the mechanical OAM operator, or 
{\it physical} nature of the mechanical OAM.

\section{Interpretation of the analysis results}
\label{sec4}

The issue of gauge choice in the Landau problem is an unexpectedly
perplex problem, which sometimes causes confusion.
Remember first the wide-spread traditional viewpoint as follows.
There are three typical choices of gauge in the Landau problem, i.e. the symmetric 
gauge and two Landau gauges. The choice of the symmetric gauge naturally 
leads to the Laguerre type solutions, which respect the rotational (or axial) 
symmetry around the origin. On the other hand, if one chooses either of the
two Landau gauges, one is naturally led to the Hermite type solutions,
which respects the translational symmetry along the $x$-axis or the
$y$-axis. We think that this wide-spread understanding is nothing wrong
and it offers an established point of view on the issue of gauge choice 
in the Landau problem.
In another viewpoint, however, the connection
between the choice of the gauge potential and the choice of the types of
the Landau wave functions is not necessarily 
mandatory \cite{Haugset1993},\cite{Govaerts2009}.
In fact, as we have seen, independently of the choice of the gauge 
potential, there exist three conserved quantities in the Landau problem.
They are the conserved OAM $L^{cons}_z$ and the two conserved
momenta $p^{cons}_x$ and $p^{cons}_y$.
The eigen-functions of the Landau problem can be obtained by simultaneously 
diagonalizing the Landau Hamiltonian and one of the above three
conserved operators.
If we choose $L^{cons}_z$ there, the eigen-functions are characterized
by two quantum numbers, $n$ and $m$, where $n$ is the familiar Landau 
quantum number and $m$ is the eigen-value of the OAM operator
$L^{cons}_z$. We emphasize once again that this is true for
any choice of the gauge potential configuration. 
Naturally, the simplest candidate of this type
of solutions is the familiar eigen-states 
$\vert \,\Psi^{(S)}_{n,m} \rangle$ obtained with the choice of the symmetric
gauge potential $\bm{A}^{(S)}$.
However, let us consider such states that are obtained from 
the symmetric gauge eigen-states $\vert \,\Psi^{(S)}_{n,m} \rangle$ by
operating a gauge transformation matrix $U^{(\chi)} = e^{\,- \,i \,e \,\chi (\bm{r})}$
as
\begin{equation}
 \vert \,\Psi^{(\chi)} \rangle \ = \ U^{(\chi)} \,\vert \,\Psi^{(S)}_{n,m} \rangle ,
 \label{Eq:Psi_chi}
\end{equation}
where $\chi (\bm{r})$ is assumed to be an arbitrary harmonic function.
Then, it seems obvious that the new states $\vert \,\Psi^{(\chi)} \rangle$
are also characterized by the two quantum numbers $n$ and $m$.
Since the dependence on the Landau quantum number needs little
explanation, here we check the dependence on the second quantum
number $m$, just to be sure. Operating $L^{cons}_z (\bm{A}^{(\chi)})$
on $\vert \,\Psi^{(\chi)} \rangle$, we obtain
\begin{eqnarray}
 L^{cons}_z (\bm{A}^{(\chi)}) \,\vert \,\Psi^{(\chi)} \rangle \ &=& \ 
 L^{cons}_z (\bm{A}^{(\chi)}) \,U^{(\chi)} \,\vert \,\Psi^{(S)}_{n,m} \rangle \nonumber \\
 \ &=& \ 
 U^{(\chi)} \,U^{(\chi)\dagger} \,L^{cons}_z (\bm{A}^{(\chi)}) \,
 U^{(\chi)} \,\vert \,\Psi^{(S)} \rangle \ \ \ \ \ \ \nonumber \\
 \ &=& \ U^{(\chi)} \,L^{cons}_z (\bm{A}^{(S)}) \,\vert \,\Psi^{(S)}_{n,m} \rangle 
 \nonumber \\
 \ &=& \ 
 m \,U^{(\chi)} \,\vert \,\Psi^{(S)}_{n,m} \rangle . \ \ \ \ \  
\end{eqnarray}
Here, use has been made of the
{\it covariant} gauge-transformation property of $L^{cons}_z$,
i.e. the relation $U^{(\chi) \dagger} \,L^{cons}_z (\bm{A}^{(\chi)}) \,U^{(\chi)} = 
L^{cons}_z (\bm{A}^{(S)})$.
The above equation shows that $\vert \,\Psi^{(\chi)} \rangle$ are the eigen-functions
of $L^{cons}_z$ with the eigen-value $m$.
We thus confirm that the states $\vert \,\Psi^{(\chi)} \rangle$ are in fact
characterized by the quantum numbers $n$ and $m$, so that it is
legitimate to write (\ref{Eq:Psi_chi}) as $\vert \,\Psi^{(\chi)}_{n,m} \rangle =  
U^{(\chi)} \,\vert \,\Psi^{(S)}_{n,m} \rangle$.
Note that the eigen-states $\vert \,\Psi^{(\chi)}_{n,m} \rangle$ obtained with 
arbitrary (regular) gauge function $\chi$ are different from the symmetric-gauge
eigen-states only by the phase factor $e^{\,- \,i \,e \,\chi}$, so that
the electron probability densities corresponding to $\vert \,\Psi^{(\chi)}_{n,m} \rangle$
and $\vert \,\Psi^{(S)}_{n,m} \rangle$ must be exactly the same and they show
the axial symmetry around the coordinate origin.
Thus, the set of eigen-functions $\vert \,\Psi^{(\chi)}_{n,m} \rangle$ obtained
in this way may be called the gauge-potential-independent extensions of the
symmetric gauge eigen-states $\vert \,\Psi^{(S)}_{n,m} \rangle$.
(Alternatively, mimicking the terminology advocated in the recent literature
\cite{Lorce2013},\cite{Review_LL2014}, they might simply be called the 
gauge-invariant extension based on the symmetric-gauge eigen-states 
$\vert \,\Psi^{(S)}_{n,m} \rangle$.)  

Exactly by the same logic, we can define the states, which may be called
the gauge-potential-independent extension of the 1st Landau-gauge 
eigen-states $\vert \,\Psi^{(L_1)}_{n, k_x} \rangle$ by
\begin{equation}
 \vert \,\Psi^{(\chi)}_{n,k_x} \rangle \ = \ U^{(\chi)} \,\vert \,\Psi^{(L_1)}_{n, k_x} \rangle .
\end{equation}
Here, $U^{(\chi)} = e^{\,- \,i \,e \,\chi}$ with $\chi$ being an arbitrary 
harmonic function.
Naturally, any of the eigen-states $\vert \,\Psi^{(\chi)}_{n, k_x} \rangle$
have exactly the same electron probability densities as the 1st Landau-gauge
eigen-states $\vert \,\Psi^{(L_1)}_{n, k_x} \rangle$. 
However, it is also obvious that these densities are absolutely different from 
the probability densities corresponding to the symmetric-gauge eigen-states
$\vert \,\Psi^{(S)}_{n,m} \rangle$. Undoubtedly,
it is related to the fact that there is no gauge transformation which directly 
connects the eigen-states of the symmetric gauge and those of the 1st
Landau gauge. 
The reason may further be traced back to
the (infinitely-many) degeneracy of the Landau levels, which happens in both 
of the symmetric-gauge eigen-states and of the Landau-gauge eigenstates.  
To understand this state of affairs in a more concrete manner, 
suppose that we operate the gauge
transformation matrix $U = e^{\,i \,\frac{1}{2} \,e \,B \, x \,y}$ on the
symmetric-gauge eigen-states $\vert \,\Psi^{(S)}_{n,m} \rangle$. 
Then, using the completeness relation for the 1st Landau-gauge 
eigen-states within the Hilbert space of fixed number of the Landau quantum 
number $n$ given as
\begin{equation}
 \int \,d k_x \,\,\vert \,\Psi^{(L_1)}_{n, k_x} \rangle \,\langle \Psi^{(L_1)}_{n, k_x} \,\vert
 \ = \ 1,
\end{equation}
we obtain
\begin{eqnarray}
 U \,\vert \,\Psi^{(S)}_{n,m} \rangle \ &=& \ \int \,d k_x \,\,
 \vert \,\Psi^{(L_1)}_{n, k_x} \rangle \,\langle \Psi^{(L_1)}_{n, k_x} \,\vert  \,U \,
 \vert \,\Psi^{(S)}_{n,m} \rangle \nonumber \\
 &=& 
 \int \,d k_x \,\,U_{n, k_x \,;\, n,m} \,\vert \,\Psi^{(L_1)}_{n, k_x} \rangle ,
\end{eqnarray}
with the definition
\begin{eqnarray}
 U_{n, k_x \,;\, n,m} \ \equiv \ \langle \Psi^{(L_1)}_{n, k_x} \,\vert \,U \,\vert \,
 \Psi^{(S)}_{n,m} \rangle .
\end{eqnarray}
This means that the gauge-transformed states $U \,\vert \,\Psi^{(S)}_{n,m} \rangle$
are superpositions of the 1st Landau-gauge eigen-states
$\vert \,\Psi^{(L_1)}_{n, k_x} \rangle$ with the weight function
$U_{n, k_x \,;\, n, m}$. The explicit form of this weight function is already
written down in some previous literature \cite{Haugset1993},\cite{WKZ2018}. 
It is given as
\begin{equation}
 U_{n, k_x \,;\, n,m} \ = \ C_{n,m} \,H_{n-m} \left( \frac{y_0}{l_B} \right) \,
 e^{\,- \,\frac{y^2_0}{2 \,l^2_B}} ,
\end{equation}
with $y_0 = k_x \,/ (e \,B)$ and
\begin{equation}
 C_{n,m} \ = \ l_B \,\left( \frac{1}{\sqrt{\pi} \,2^{n-m} \,(n - m) \,! \,l_B} \right)^{1/2}. 
\end{equation}
Here, one may notice that the states obtained by operating 
$U = e^{\,i \,\frac{1}{2} \,e \,B \,x \,y}$ on $\vert \,\Psi^{(S)}_{n,m} \rangle$ are
nothing but the states $\vert \,\Psi^{(L_1)}_{n,m} \rangle$ defined before 
by Eqs.(\ref{Eq:Psi_L1_n_m_A}) and (\ref{Eq:Psi_L1_n_m_B}).
From this fact, we now realize that
the states $\vert \,\Psi^{(L_1)}_{n,m} \rangle$ introduced there are actually 
the following superposition of the 1st Landau-gauge eigen-states
$\vert \Psi^{(L_1)}_{n, k_x} \rangle$, 
\begin{equation}
 \vert \,\Psi^{(L_1)}_{n,m} \rangle \ = \ \int \,d k_x \,\,
 U_{n, k_x \,;\, n, m} \,\vert \,\Psi^{(L_1)}_{n, k_x} \rangle .
 \label{Eq:trans}
\end{equation}
Obviously, this relation can be generalized to arbitrary states
defined as $\vert \,\Psi^{(\chi)}_{n,m} \rangle = U^{(\chi)} \,\vert \,\Psi^{(S)}_{n,m} \rangle
= e^{\,- i \,\,e \,\chi} \,\vert \,\Psi^{(S)}_{n,m} \rangle$ with 
$\chi$ being an arbitrary harmonic
function, and they are also the members of the gauge-potential-independent 
extension of the symmetric-gauge eigen-states $\vert \,\Psi^{(S)}_{n,m} \rangle$.
In this way, we are led to the conclusion that there are totally three such 
extensions, i.e. the gauge-potential-independent extension based on the 
symmetric gauge eigen-states, that based on the 1st Landau-gauge 
eigen-states and that 
based on the 2nd Landau-gauge eigen-states.
What is important to recognize here is that these three types of
eigen-states belong to totally different (or inequivalent) gauge classes.
 
What is meant by the above statement would be understood by checking 
the following two properties : 
\begin{itemize}
\item The expectation values of a genuinely {\it gauge-invariant} physical quantity
should be the {\it same} irrespectively of the choice of three types of
eigen-functions.

\item The expectation values of the {\it gauge-variant} physical quantity
can be {\it different} for the eigen-states belonging to different gauge classes.
\end{itemize}

Through the analyses in the previous section, we have verified that
the expectation value of the canonical momentum operator $p^{can}_x$ and 
the canonical OAM operator $L^{can}_z$ are in fact different between the
two states belonging to different gauge classes, i.e. between the 
gauge-potential-independent
extension based on the symmetric gauge eigen-states and that based on the 
1st Landau gauge eigen-states. 
This is not surprising at all, since these canonical quantities
are widely believed to be gauge-variant ones.
In sharp contrast, we found that the expectation value of the mechanical
momentum operator $p^{mech}_x$ and the mechanical OAM operator
$L^{mech}_z$ perfectly coincide between the two states belonging to
different gauge classes.
Somewhat perplexing are the conserved momentum operator $p^{cons}_x$
and the conserved OAM operator $L^{cons}_z$.
Despite the fact that these two operators transform in a gauge-covariant
manner just like the mechanical operators, we found that the expectation values of 
these conserved operators do not coincide between the eigen-states belonging to 
two gauge-inequivalent classes. What is the cause of this remarkable difference ?

It can be understood as follows. Notice first that the 
following equality holds :
\begin{eqnarray}
 \sum_m \,U_{n, k^\prime_x ; n, m} \,U^*_{n, k_x ; n, m} &=&
 \sum_m \langle \Psi^{(L_1)}_{n, k^\prime_x} \,\vert \,U \,
 \vert \,\Psi^{(S)}_{n, m} \rangle \,\langle \Psi^{(S)}_{n,m} \,\vert \,U^\dagger \,\vert \,
 \Psi^{(L_1)}_{n, k_x} \rangle \hspace{8mm} \nonumber \\
 &=& \langle \Psi^{(L_1)}_{n, k^\prime_x} \,\vert \,U \,U^\dagger \,\vert \,
 \Psi^{(L_1)}_{n, k_x} \rangle \ = \ \delta (k^\prime_x - k_x) .
\end{eqnarray}
By using this relation, (\ref{Eq:trans}) can be inverted as 
\begin{equation}
 \vert \,\Psi^{(L_1)}_{n, k_x} \rangle \ = \ \sum_m \,U^*_{n, k_x ; n, m} \,\vert \,
 \Psi^{(L_1)}_{n, m} \rangle .
\end{equation}
Now we compare the matrix elements of the mechanical OAM operator
and the conserved OAM operator between the states
$\vert \,\Psi^{(L_1)}_{n, k^\prime_x} \rangle$ and
$\vert \,\Psi^{(L_1)}_{n, k_x} \rangle$.
For the former quantity, we find that
\begin{eqnarray}
 &\,&
 \langle \Psi^{(L_1)}_{n, k^\prime_x} \,\vert \,\hat{L}^{mech}_z (\bm{A}^{(L_1)}) \,
 \vert \,\Psi^{(L_1)}_{n, k_x} \rangle \nonumber \\
 &=&
 \sum_{m^\prime, m} \,U_{n, k^\prime_x ; n, m^\prime} \,
 U^*_{n, k_x ; n, m} \,\,\langle \Psi^{(L_1)}_{n, m^\prime} \,\vert \,
 \hat{L}^{mech}_z (\bm{A}^{(L_1)}) \,\vert \,\Psi^{(L_1)}_{n, m} \rangle \hspace{6mm}
 \nonumber \\
 &=& \sum_{m^\prime, m} \,U_{n, k^\prime_x ; n, m^\prime} \,
 U^*_{n, k_x ; n, m} \times (2 \,n + 1) \,\delta_{m^\prime, m} \nonumber \\
 &=& (2 \,n + 1) \,\sum_{m^\prime, m} \,U_{n, k^\prime_x ; n, m^\prime} \,
 U^*_{n, k_x ; n, m} \nonumber \\
 &=& \ (2 \,n + 1) \,\delta (k^\prime_x - k_x) ,
\end{eqnarray}
which should be compared with the matrix elements between the symmetric-gauge
eigen-states given as $\langle \Psi^{(S)}_{n,m^\prime} \,\vert \,
\hat{L}^{mech}_z (\bm{A}^{(S)}) \,\vert \,\Psi^{(S)}_{n,m} \rangle = 
(2 \,n + 1) \,\delta_{m^\prime, m}$.
A noteworthy fact here is that the Landau quantum number $n$ appears as a
common quantum number in the 1st Landau-gauge eigen-states and the
symmetric gauge ones and that, except for the appearance of the delta function
and the Kronecker delta related to the different normalization conditions, the 
matrix elements of the mechanical OAM operators in both types of eigen-states 
depend only on $n$ and they are just the same, which may be interpreted as showing
the physical nature of the mechanical OAM. 

In contrast, for the matrix elements of the conserved OAM operator, we have
\begin{eqnarray}
 &\,&
 \langle \Psi^{(L_1)}_{n, k^\prime_x} \,\vert \,\hat{L}^{cons}_z (\bm{A}^{(L_1)}) \,
 \vert \,\Psi^{(L_1)}_{n, k_x} \rangle \nonumber \\
 &=&
 \sum_{m^\prime, m} \,U_{n, k^\prime_x ; n, m^\prime} \,
 U^*_{n, k_x ; n, m} \,\,\langle \Psi^{(L_1)}_{n, m^\prime} \,\vert \,
 \hat{L}^{cons}_z (\bm{A}^{(L_1)}) \,\vert \,\Psi^{(L_1)}_{n, m} \rangle \hspace{6mm}
 \nonumber \\
 &=& \sum_{m^\prime, m} \,U_{n, k^\prime_x ; n, m^\prime} \,
 U^*_{n, k_x ; n, m} \times m \,\delta_{m^\prime, m} \nonumber \\
 &=& \sum_{m} \,U_{n, k^\prime_x ; n, m} \,
 U^*_{n, k_x ; n, m} \times m \nonumber \\
 &=& \left\{ \,n + \frac{1}{2} - \frac{k_x^2}{2 \,e \,B} \right\} \,
 \delta (k^\prime_x - k_x) \ + \,\frac{e \,B}{2} \,
 \delta^{\prime \prime} (k^\prime_x - k_x) ,
\end{eqnarray}
which should be compared with the matrix element between the symmetric-gauge
eigen-states given as $\langle \Psi^{(S)}_{n,m^\prime} \,\vert \,
\hat{L}^{cons}_z (\bm{A}^{S)}) \,\vert \,\Psi^{(S)}_{n,m} \rangle = m \,\delta_{m^\prime, m}$.

One should not overlook the fact that the matrix element of the 
conserved OAM operator between the
1st Landau-gauge eigen-states depend on both of $n$ and $k_x$, 
while the matrix element of the conserved OAM operator between the
symmetric-gauge eigen-states depends only on $m$. Undoubtedly,
this comparison indicates unphysical (or nonobservable) nature of the quantum 
number $k_x$ and $m$ appearing in the two different types of Landau eigen-functions
Hamiltonian. In fact, we recall that nonobservability of the quantum number 
$m$ or the canonical OAM in the Landau problem was an object of extensive
discussion in recent papers \cite{WKZZ2020},\cite{WKZZ2021}. 
(Clearly, the same can be said also for the quantum-number 
$k_x$ or the canonical momentum.).

In any case, the above comprehensive analysis, combined with the comparison
of Table 1 and Table 2, confirms the fact
that the gauge-independence of the expectation value
of the conserved OAM operator is restricted within the {\it same} gauge class 
and it cannot be extended to {\it different} or {\it inequivalent} gauge classes.
In other words, the {\it gauge-covariance} of a certain operator does not
necessarily mean the {\it gauge-independence} of the corresponding quantity.
We emphasize that this is a highly nontrivial finding which has never been 
explicitly stated before at least in an analytically solvable model like 
the Landau problem.

Before ending this section, we think it enlightening as well as important
to point out an interesting
relationship between the conserved OAM discussed in this paper and the
idea of gauge-invariant (or more precisely gauge-covariant) extension
of the canonical OAM advocated in the literature on the nucleon spin
decomposition problem. The latter quantity was introduced based on
the idea of Chen et al. \cite{Chen2008},\cite{Chen2009}, 
who proposed the decomposition of
the gauge field into the {\it physical} component and the {\it pure-gauge}
component as
\begin{equation}
 \bm{A} \ = \ \bm{A}^{phys} \ + \ \bm{A}^{pure} .
\end{equation}
The basic postulate of theirs is that, under a gauge transformation specified
by $U = e^{\,i \,e \,\chi}$, the above two components transform as
\begin{eqnarray}
 \bm{A}^{phys} &\rightarrow& \bm{A}^{\prime \,phys} \ = \ \bm{A}^{phys}, \\
 \bm{A}^{pure} &\rightarrow& \bm{A}^{\prime \,pure} \ = \ 
 \bm{A}^{pure} \ + \ \nabla \chi .
\end{eqnarray}
That is, the gauge degrees of freedom are totally carried by the
pure-gauge part $\bm{A}^{pure}$, while the physical component 
$\bm{A}^{phys}$ is intact under a gauge transformation.

Suppose for the moment that such a decomposition in fact exists. 
Then, one may introduce the following quantities : 
\begin{eqnarray}
 p^{g.c.c. [\bm{A}^{phys}]}_x \ &\equiv& \ p^{can}_x \ + \ e \,( A_x - A^{phys}_x )
 \nonumber \\
 &=& p^{mech}_x \ - \ e \,A^{phys}_x , \\
 L^{g.c.c. [\bm{A}^{phys}]}_z \ &\equiv& \ L^{can}_z \ + \ e \,
 \left[ \bm{r} \times (\bm{A} - \bm{A}^{phys} ) \right]_z \nonumber \\
 &=& 
 L^{mech}_z \ - \ e \,( \bm{r} \times \bm{A}^{phys})_z ,
\end{eqnarray}
which may be called the gauge-covariant-canonical (g.c.c.) momentum
and  the g.c.c OAM.
(In the previous literature \cite{Review_LL2014}, 
they were called the gauge-invariant-canonical (g.i.c.) 
momentum and  g.i.c OAM. However, in the context of quantum mechanics
or quantum field theory, it would be more legitimate to use the word 
{\it covariant} rather than the word {\it invariant}.)
In fact, it is obvious that these operators transform covariantly
under an arbitrary gauge transformation, because the mechanical 
operators transform covariantly, while $\bm{A}^{phys}$ is intact.
A pitfall of the above argument is that the decomposition of the
vector potential into the physical and pure-gauge potential is not
always unique. Remember that, when Chen et al. proposed the above
decomposition of the gauge field, what was in their mind was the familiar 
transverse-longitudinal decomposition of the vector 
potential \cite{Cohen-Tannoudji1989}. 
Once the Lorentz frame of reference is fixed, 
the transverse-longitudinal decomposition is known 
to be unique as long as the condition for the Helmholtz theorem
is satisfied \cite{Zangwill2013}. Unfortunately, in our 2-dimensional Landau problem,
in which the condition for the Helmholtz theorem is not satisfied,
there is no way to uniquely fix the physical component of the
vector potential. This is reflected by the fact that the
three practical gauge choices in the Landau problem all satisfy the
{\it transverse condition} $\nabla \cdot \bm{A}^{(S)} =
\nabla \cdot \bm{A}^{(L_1)} = \nabla \cdot \bm{A}^{(L_2)} = 0$.
Thus, although a unique identification of the physical component
is not possible, suppose that we asign by hand the symmetric gauge potential
as the physical component, i.e. $\bm{A}^{(S)} \equiv \bm{A}^{phys}$.
Then, it follows that
\begin{eqnarray}
 p^{g.c.c. [\bm{A}^{(S)}]}_x &=& \ p^{can}_x \ + \ 
 e \,A_x \ + \ \frac{1}{2} \,e \,B \,y , \\
 L^{g.c.c. [\bm{A}^{(S)}]}_z &=& \ L^{can}_z \ + \ e \,(\bm{r} \times \bm{A})_z
 \ - \ \frac{1}{2} \,e \,B \,r^2.
\end{eqnarray}
Here, one may notice that $L^{g.c.c. [\bm{A}^{(S)}]}_z$ above precisely 
coincides with our conserved OAM, i.e.
\begin{equation}
 L^{g.c.c. [\bm{A}^{(S)}]}_z \ = \ L^{cons}_z.
\end{equation}

\vspace{2mm} 
On the other hand, if the 1st Landau gauge potential is identified with
the physical component as $\bm{A}^{(L_1)} \equiv \bm{A}^{phys}$, we have
\begin{eqnarray}
 p^{g.c.c. [\bm{A}^{(L_1)}]}_x &=& \ p^{can}_x \ + \ e \,A_x \ + \ e \,B \,y ,\\
 L^{g.c.c. [\bm{A}^{(L_1)}]}_z &=& \ L^{can}_z \ + \ e \,(\bm{r} \times \bm{A})_z
 \ - \ e \,B \,y^2 .
\end{eqnarray}
Here, one finds that $p^{g.c.c. [\bm{A}^{(L_1)}]}_x$ precisely coincides with our
conserved momentum, i.e.
\begin{equation}
 p^{g.c.c. [\bm{A}^{(L_1)}]}_x \ = \ p^{cons}_x .
\end{equation}
What can we learn from the above consideration ?
We have already shown that the conserved momentum as well as
the conserved OAM are not truly gauge-invariant quantities despite their
gauge-covariant transformation property.
Undoubtedly, the same can be said for the gauge-covariant extension
of the canonical OAM advocated in the recent literature \cite{Hatta2011}
\nocite{Lorce2011}-\cite{Review_LL2014}.  

The argument above also reminds us of the debates on the gauge-invariant
or gauge-variant nature of the gluon spin as well as the canonical
OAM of quarks in the nucleon spin decomposition problem.
Originally, the canonical quark OAM inside the nucleon appearing
in the famous Jaffe-Manohar decomposition was believed to be a 
{\it gauge-variant} quantity \cite{JM1990}.
However, after Chen et al.'s paper appeared \cite{Chen2008}\cite{Chen2009}, 
several authors proposed a concept of gauge-covariant extension of the
canonical OAM operator and the belief, that this extended 
canonical OAM can be thought of as a gauge-invariant quantity,
became popular. (See, for example, the review \cite{Review_LL2014}.)
However, this extension of the canonical OAM needs the
concept of {\it physical component} of the gauge field.
As we have clearly shown in the present paper, when there exist plural
possibilities for the choice of the physical component of the
gauge field, such extensions of the canonical OAM actually
depends on the basis gauge of extensions and they are not
gauge-invariant quantities in the rigorous sense.
Note that the situation is totally different for the mechanical quark
OAM inside the nucleon, which is defined without need
of the physical component of the gauge field.
One can then say that the mechanical OAM is a genuinely
gauge-invariant quantity \cite{Ji1997}\nocite{Waka2010}-\cite{Waka2011}.
Unfortunately, to make the gluon spin inside the nucleon
gauge-invariant, we also need the concept of physical
component of the gluon field and the gauge-invariant
extension based on it. Although the gauge-invariant extension
based on the light-cone gauge is practically the most useful choice
in the deep-inelastic-scattering physics, still one should keep in
mind the fact that the gauge-invariance of the gluon
spin attained in that way is not a gauge-invariance in the true sense.
Undoubtedly, this obstraction must have a deep connection with 
the long-known fact in the field of pertubative QCD that there is no 
{\it local} and {\it gauge-invariant} gluon spin operator.

\section{Conclusion}
\label{sec5}

In the present paper, we proposed a simple quantum mechanical formulation
of the famous Landau problem, which enables us to avoid a
specific choice of gauge potential when writing down the eigen-functions of the
Landau problem. The formalism is based on the 
existence of three conserved quantities in the Landau problem, i.e. the two 
conserved momenta and one conserved OAM, absolutely independently of 
the choice of gauge potential. 
A prominent feature of the above conserved momenta and the conserved OAM,
which are also called the pseudo momenta and the pseudo OAM in some
literature, is that they have covariant tranformation properties under an arbitrary
gauge transformation, just like the familiar mechanical momentum and 
mechanical OAM operators. (The latters are widely believed to be
manifestly gauge-invariant quantities.) 
In this gauge-potential-independent formulation, although the quantum mechanical 
eigen-functions of the Landau Hamiltonian can be written down without 
fixing the gauge potential, it turns out that these solutions are divided into 
three classes, which we may call the gauge-potential-independent extensions 
based on three different basis gauges. i.e. the 1st Landau gauge, 
the 2nd Landau gauge, and the symmetric gauge. 
We have carried out a comparative analysis of the matrix elements of the
mechanical operators and the conserved operators between the three different
classes of eigenstates. We then found that the matrix elements of the
mechanical momentum and mechanical OAM operators between the three
different eigen-states perfectly coincide with each other. This is
interpreted to verify the genuinely gauge-invariant nature of the mechanical
quantities. On the other hand, it turned out that the matrix elements of
the conserved momentum and conserved OAM operators between the
three different eigen-states do not coincide with each other. 
This means that three gauge-potential-independent extensions do actually 
belong to {\it different gauge classes}, and that the conserved momentum
and conserved OAM are {\it not} truly gauge-invariant physical quantities
despite their covariant gauge-transformation property.
This also dictates that little physical meaning can be given to the idea of the 
gauge-invariant extension of the canonical OAM advocated in the recent 
literature on the nucleon spin decomposition problem.
We can also say that the present analysis provides us with one concrete and
clear example in which the gauge symmetry is just a redundancy of the 
description with no substancial physical contents \cite{Zee2010},\cite{Schwartz2014}.








\appendix



\section{Calculation of the matrix elements in the $\vert n, k_x \rangle$-basis}
\renewcommand{\theequation}{A.\arabic{equation} }
\setcounter{equation}{0}

By using the explicit form of $\Psi^{(L_1)}_{n, k_x} (x, y)$ given 
by (\ref{Eq:Eigen_function_Psi_L1_n_kx}),
it can readily be shown that
\begin{equation}
 \langle \Psi^{(L_1)}_{n, k^\prime_x} \,\vert \,\hat{p}^{can}_x \,\vert \,\Psi^{(L_1)}_{n, k_x} \rangle
 \ = \ \langle \Psi^{(L_1)}_{n, k^\prime_x} \,\vert \,- \,i \,\frac{\partial}{\partial x} \,
 \vert \,\Psi^{(L_1)}_{n, k_x} \rangle \ = \ k_x \,\delta (k^\prime_x - k_x) .
\end{equation}
Furthermore, since $\hat{p}^{cons}_x (\bm{A})$ reduces to the canonical 
momentum $\hat{p}^{can}_x$ in the $L_1$-gauge, we naturally have
\begin{equation}
 \langle \Psi^{(L_1)}_{n, k^\prime_x} \,\vert \,\hat{p}^{cons}_x (\bm{A}^{(L_1)}) \,
 \vert \,\Psi^{(L_1)}_{n, k_x} \rangle \ = \ 
 \langle \Psi^{(L_1)}_{n, k^\prime_x} \,\vert \,\hat{p}^{can}_x \,\vert \,\Psi^{(L_1)}_{n, k_x} \rangle
 \ = \ k_x \,\delta (k^\prime_x - k_x) .
\end{equation}
Next, the matrix element of the mechanical momentum operator becomes
\begin{eqnarray}
 &\,&
 \langle \Psi^{(L_1)}_{n, k^\prime_x} \,\vert \,\hat{p}^{mech}_x (\bm{A}^{(L_1)}) \,
 \vert \,\Psi^{(L_1)}_{n, k_x} \rangle \nonumber \\
 &=& 
 \langle \Psi^{(L_1)}_{n, k^\prime_x} \,\vert \,\hat{p}^{can}_x \ + \ e \,(\,- \,B \,y) \,
 \vert \,\Psi^{(L_1)}_{n, k_x} \rangle \nonumber \\
 &=& k_x \,\delta (k^\prime_x - k_x) \, - \, e \,B \,
 \delta (k^\prime_x - k_x) \,\langle Y_n \,\vert \,y \,\vert \,Y_n \rangle.
\end{eqnarray}
Here, the matrix element $\langle Y_n \,\vert \,y \,\vert \,Y_n \rangle$ can be
evaluated as follows : 
\begin{eqnarray}
 \langle Y_n \,\vert \,y \,\vert \, Y_n \rangle \ &=& \
 N^2_n \,\int_{- \,\infty}^\infty \,d y \,e^{\,- \,\frac{(y - y_0)^2}{l^2_B}} \,
 H_n \left( \frac{y - y_0}{l_B} \right) \,y \,
 H_n \left( \frac{y - y_0}{l_B} \right) \nonumber \\
 \ &=& \ y_0 \ = \ \frac{k_x}{e \,B} .
\end{eqnarray}
We therefore find a remarkable relation as
\begin{eqnarray}
 \langle \Psi^{(L_1)}_{n, k^\prime_x} \,\vert \,\hat{p}^{mech}_x (\bm{A}^{(L_1)}) \,
 \vert \,\Psi^{(L_1)}_{n, k_x} \rangle &=& 
 k_x \,\delta (k^\prime_x - k_x) \ - \ e \,B \,\frac{k_x}{e \,B} \,
 \delta (k^\prime_x - k_x) \hspace{6mm} \nonumber \\
 &=& \hspace{5mm} 0,
\end{eqnarray}
the physical significance of which is explained in the main text.

\vspace{2mm}
Next we turn to the matrix elements between the eigen-states
$\vert \,\Psi^{(S)}_{n, k_x} \rangle$. With the use of 
the relation $\vert \,\Psi^{(S)}_{n, k_x} \rangle  =  
U^\dagger \,\vert \,\Psi^{(L_1)} \rangle$, we obtain
\begin{equation}
 \langle \Psi^{(S)}_{n, k^\prime_x} \,\vert\,\hat{p}^{can}_x \,\vert \,\Psi^{(S)}_{n, k_x} \rangle
 \ = \ \langle \Psi^{(L_1)}_{n, k^\prime_x} \,\vert \,U \,\hat{p}^{can}_x \,U^\dagger \,
 \vert \,\Psi^{(L_1)}_{n, k_x} \rangle .
\end{equation} 
Noting that
\begin{equation}
 U \,\hat{p}^{can}_x \,U^\dagger \ = \ \hat{p}^{can}_x \ - \ \frac{1}{2} \,e \,B \,y ,
\end{equation}
we therefore find that
\begin{eqnarray}
 &\,& \langle \Psi^{(S)}_{n, k^\prime_x} \,\vert \,\hat{p}^{can}_x \,\vert \,\Psi^{(S)}_{n, k_x} \rangle
 \ = \  \langle \Psi^{(L_1)}_{n, k^\prime_x} \,\vert \,
 \hat{p}^{can}_x \ - \ \frac{1}{2} \,e \,B \,y \,\vert \,\Psi^{(L_1)}_{n, k_x} \rangle \nonumber \\
 &\,& = k_x \,\delta (k^\prime_x - k_x) \ - \ \frac{1}{2} \,e \,B \,\,\frac{k_x}{e \,B} \,
 \delta (k^\prime_x - k_x) \ = \ \frac{1}{2} \,k_x \,\delta (k^\prime_x - k_x) . \ \ \ \ \ 
\end{eqnarray}
Similarly, we have
\begin{equation}
 \langle \Psi^{(S)}_{n, k^\prime_x} \,\vert \,\hat{p}^{cons}_x (\bm{A}^{(S)}) \,
 \vert \,\Psi^{(S)}_{n, k_x} \rangle \ = \ 
 \langle \Psi^{(L_1)}_{n, k^\prime_x} \,\vert \,U \,\hat{p}^{cons}_x (\bm{A}^{(S)}) \,U^\dagger \,
 \vert \,\Psi^{(L_1)}_{n, k_x} \rangle .
\end{equation}
Taking care of the covariant gauge-transformation property of $\hat{p}^{cons}_x$,
which means the relation
$U \,\hat{p}^{cons}_x (\bm{A}^{(S)}) \,U^\dagger \ = \ \hat{p}^{cons}_x (\bm{A}^{(L_1)})$,
we therefore find that
\begin{eqnarray}
 \langle \Psi^{(S)}_{n, k^\prime_x} \,\vert \,\hat{p}^{cons}_x (\bm{A}^{(S)}) \,
 \vert \,\Psi^{(S)}_{n, k_x} \rangle &=& 
 \langle \Psi^{(L_1)}_{n, k^\prime_x} \,\vert \,\hat{p}^{cons}_x (\bm{A}^{(L_1)}) \,
 \vert \,\Psi^{(L_1)}_{n, k_x} \rangle \nonumber \\
 &=& k_x \,\delta (k^\prime_x - k_x) .
\end{eqnarray} 
For the matrix element of the mechanical momentum operator, we obtain
\begin{eqnarray}
 &\,&
 \langle \Psi^{(S)}_{n, k^\prime_x} \,\vert\,\hat{p}^{mech}_x (\bm{A}^{(S)}) \,
 \vert \,\Psi^{(S)}_{n, k_x} \rangle \ = \
 \langle \Psi^{(L_1)}_{n, k^\prime_x} \,\vert \,U \,\hat{p}^{mech}_x (\bm{A}^{(S)}) \,
 U^\dagger \vert \,\Psi^{(S)}_{n, k_x} \rangle \hspace{10mm} \nonumber \\
 &=& \langle \Psi^{(L_1)}_{n, k^\prime_x} \,\vert \,\hat{p}^{mech}_x (\bm{A}^{(L_1)}) \,
 \vert \,\Psi^{(L_1)}_{n, k_x} \rangle \ = \  
 \langle \Psi^{(L_1)}_{n, k^\prime_x} \,\vert \,\hat{p}^{can}_x \ - \ e \,B \,y \,
 \vert \,\Psi^{(L_1)}_{n, k_x} \rangle \nonumber \\ 
 &=& \ k_x \,\delta (k^\prime_x - k_x) \ - \ e \,B \,\,\frac{1}{e \,B} \,\,
 k_x \,\delta (k^\prime_x - k_x) \ = \ 0.
\end{eqnarray}
Here, we have used the fact that $\hat{p}^{mech}_x$ also transforms
covariantly under a gauge transformation.

Next, we evaluate the matrix elements of the three OAM operators
in the two $\vert \,n, k_x \rangle$-basis functions, 
$\Psi^{(L_1)}_{n, k_x} (x,y)$ and $\Psi^{(S)}_{n, k_x} (x,y)$.
Although it may sound strange, this has never been done before.
The reason is probably because the symmetric gauge eigen-states 
$\vert \,\Psi^{(S)}_{n,m} \rangle$ are the most natural and convenient basis
to deal with the OAM operators, and one seldom paid attention to
calculating the expectation values of the OAM operators between the
eigen-functions $\vert \,\Psi^{(L_1)}_{n, k_x} \rangle$ in the 1st Landau
gauge.

With the use of the definitions of $\hat{L}^{cons}_z (\bm{A})$ and 
$\hat{L}^{mech}_z (\bm{A})$, we obtain
\begin{eqnarray}
 &\,&
 \langle \Psi^{(L_1)}_{n, k^\prime_x} \,\vert \, \hat{L}^{cons}_z (\bm{A}^{(L_1)}) \,
 \vert \,\Psi^{(L_1)}_{n, k_x} \rangle \ = \   
 \langle \Psi^{(L_1)}_{n, k^\prime_x} \,\vert \, \hat{L}^{can}_z \,
 \vert \,\Psi^{(L_1)}_{n, k_x} \rangle \nonumber \\
 &\,& \hspace{15mm} \ - \ \frac{1}{2} \,e \,B \,
 \langle \Psi^{(L_1)}_{n, k^\prime_x} \,\vert \, x^2 \,
 \vert \,\Psi^{(L_1)}_{n, k_x} \rangle \, + \, \frac{1}{2} \,e \,B \,
 \langle \Psi^{(L_1)}_{n, k^\prime_x} \,\vert \, y^2  \,
 \vert \,\Psi^{(L_1)}_{n, k_x} \rangle , \ \ \ \ \ \\
 &\,&
 \langle \Psi^{(L_1)}_{n, k^\prime_x} \,\vert \, \hat{L}^{mech}_z (\bm{A}^{(L_1)}) \,
 \vert \,\Psi^{(L_1)}_{n, k_x} \rangle \nonumber \\ 
 &\,& \hspace{15mm} \ = \ 
 \langle \Psi^{(L_1)}_{n, k^\prime_x} \,\vert \, \hat{L}^{can}_z \,
 \vert \,\Psi^{(L_1)}_{n, k_x} \rangle \, + \, 
 e \,B \,
 \langle \Psi^{(L_1)}_{n, k^\prime_x} \,\vert \, y^2  \,
 \vert \,\Psi^{(L_1)}_{n, k_x} \rangle ,
\end{eqnarray}
Similarly, using the relation $\vert \,\Psi^{(S)}_{n, k_x} \rangle = U^\dagger \,
\vert \,\Psi^{(L_1)}_{n, k_x} \rangle$ together with the gauge-covariant 
transformation properties of $\hat{L}^{cons}_z (\bm{A})$ and 
$\hat{L}^{mech}_z (\bm{A})$, we obtain
\begin{eqnarray}
 &\,&
 \langle \Psi^{(S)}_{n, k^\prime_x} \,\vert \, \hat{L}^{cons}_z (\bm{A}^{(S)}) \,
 \vert \,\Psi^{(S)}_{n, k_x} \rangle \ = \  
 \langle \Psi^{(L_1)}_{n, k^\prime_x} \,\vert \, \hat{L}^{can}_z \,
 \vert \,\Psi^{(L_1)}_{n, k_x} \rangle \nonumber \\
 &\,& \hspace{15mm} \ - \ \frac{1}{2} \,e \,B \,
 \langle \Psi^{(L_1)}_{n, k^\prime_x} \,\vert \, x^2 \,
 \vert \,\Psi^{(L_1)}_{n, k_x} \rangle \, + \, \frac{1}{2} \,e \,B \,
 \langle \Psi^{(L_1)}_{n, k^\prime_x} \,\vert \, y^2  \,
 \vert \,\Psi^{(L_1)}_{n, k_x} \rangle , \ \ \ \ \ \\
 &\,& 
 \langle \Psi^{(S)}_{n, k^\prime_x} \,\vert \, \hat{L}^{mech}_z (\bm{A}^{(S)}) \,
 \vert \,\Psi^{(S)}_{n, k_x} \rangle \nonumber \\
 &\,& \hspace{15mm} \ = \   
 \langle \Psi^{(L_1)}_{n, k^\prime_x} \,\vert \, \hat{L}^{can}_z \,
 \vert \,\Psi^{(L_1)}_{n, k_x} \rangle \, + \, 
 e \,B \,
 \langle \Psi^{(L_1)}_{n, k^\prime_x} \,\vert \, y^2  \,
 \vert \,\Psi^{(L_1)}_{n, k_x} \rangle .
\end{eqnarray}
Thus, for evaluating the matrix elements of the three OAM operators in the
$\vert \,n, k_x \rangle$-basis, we have only to know the following three 
matrix elements :
\begin{equation}
 \langle \Psi^{(L_1)}_{n, k^\prime_x} \,\vert \, \hat{L}^{can}_z \,
 \vert \,\Psi^{(L_1)}_{n, k_x} \rangle, \ \ \  
 \langle \Psi^{(L_1)}_{n, k^\prime_x} \,\vert \, x^2 \,
 \vert \,\Psi^{(L_1)}_{n, k_x} \rangle, \ \ \  
 \langle \Psi^{(L_1)}_{n, k^\prime_x} \,\vert \, y^2 \,
 \vert \,\Psi^{(L_1)}_{n, k_x} \rangle .
\end{equation}
As one can easily verify, the two of these matrix elements, i.e.
$\langle \Psi^{(L_1)}_{n, k^\prime_x} \,\vert \, \hat{L}^{can}_z \,
\vert \,\Psi^{(L_1)}_{n, k_x} \rangle$ and
$\langle \Psi^{(L_1)}_{n, k^\prime_x} \,\vert \, y^2 \,
\vert \,\Psi^{(L_1)}_{n, k_x} \rangle$ can be calculated without much
difficulty. On the other hand, the calculation of the second 
matrix element $\langle \Psi^{(L_1)}_{n, k^\prime_x} \,\vert \, x^2 \,
\vert \,\Psi^{(L_1)}_{n, k_x} \rangle$ needs some care owing to the
plane-wave nature of the eigen-states $\vert \,\Psi^{(L_1)}_{n, k_x} \rangle$
along the $x$-direction. 
(Note that this cumbersome term appears in the matrix elements of
$\hat{L}^{can}_z$ and $\hat{L}^{cons}_z (\bm{A})$, but it does not
in those of $\hat{L}^{mech}_z (\bm{A})$.)
First, note that the eigen-functions
$\Psi^{(L_1)}_{n, k_x} (x,y)$ are normalized as
\begin{eqnarray}
 &\,&
 \int \,d x \,d y \,\Psi^{(L_1) *}_{n, k^\prime_x} (x,y) \,\Psi^{(L_1)}_{n, k_x} (x,y)
 \nonumber \\
 &=& \frac{1}{2 \,\pi} \,\int_{-\,\infty}^\infty \,d x \,
 e^{\,- \, i \,(k^\prime_x - k_x) \,x} \,
 \int_{- \,\infty}^\infty \,d y \,[Y_n (y)]^2 \ = \ 
 \delta (k^\prime_x - k_x) .
\end{eqnarray}
Here, we have used the fact that $Y_n (y)$ is normalized as 
$\int_{- \,\infty}^\infty \,d y \,[Y_n (y)]^2 = 1$.
For later convenience, we introduce the dimensionless wave function
$\psi_n (\xi)$ by
\begin{equation}
 Y_n (y) \ \equiv \ \frac{1}{\sqrt{l_B}} \,\,\psi_n (\xi) ,
\end{equation}
with $(y - y_0) \,/\,l_B = \xi$. The normalization of $\psi (\xi)$ is
then given by
\begin{equation}
 \int_{- \,\infty}^\infty \, d \xi \,[\psi_n (\xi)]^2 \ = \ 1.
\end{equation}
We begin with the calculation of the matrix element of the canonical
OAM operator : 
\begin{equation}
 \langle \Psi^{(L_1)}_{n, k^\prime_x} \,\vert \, \hat{L}^{can}_z \,\vert \,
 \Psi^{(L_1)}_{n, k_x} \rangle \ = \ 
 - \, i \,\,\langle \Psi^{(L_1)}_{n, k^\prime_x} \,\vert \, x \,\frac{\partial}{\partial y}
 \ - \ y \,\frac{\partial}{\partial x} \,\vert \,
 \Psi^{(L_1)}_{n, k_x} \rangle . \label{AppEq:L^can}
\end{equation}
Let us first consider the first part in the r.h.s.
Using the variables $\xi = (y - l^2_B \,k_x) \,/\,l_B$ and
$\xi^\prime = (y - l^2_B \,k^\prime_x) \,/\,l_B$ together with the relation
$\frac{\partial}{\partial y} = \frac{1}{l_B} \,\frac{\partial}{\partial \xi}$,
we obtain
\begin{eqnarray}
 &\,& \hspace{-5mm}
 - \,i \,\,\langle \Psi^{(L_1)}_{n, k^\prime_x} \,\vert \, x \,\frac{\partial}{\partial y} \,
 \vert \,\Psi^{(L_1)}_{n, k_x} \rangle \nonumber \\
 \ &=& \
 - \,i \,\left\{ \frac{1}{2 \,\pi} \,\int_{- \,\infty}^\infty \,d x \,
 x \,e^{\,- \,i \,(k^\prime_x - k_x) \,x} \,\right\} \,\,\frac{1}{l_B} \,\,
 \int_{- \,\infty}^\infty \,d \xi \,\psi_n (\xi^\prime) \,
 \frac{\partial}{\partial \xi} \,\psi_n (\xi) \ \ \ \ \ \nonumber \\
 \ &=& \ \left\{ - \,\frac{\partial}{\partial k_x} \,\delta (k^\prime_x - k_x) \,\right\} \,
 \frac{1}{l_B} \,\int_{- \,\infty}^\infty \, d \xi \,\psi_n (\xi^\prime) \,
 \frac{\partial}{\partial \xi} \,\psi_n (\xi) . 
\end{eqnarray}
Here, we can make use of the identity of Dirac's delta function : 
\begin{equation}
 f (k) \,\delta^\prime (k) \ = \ - \,f^\prime (0) \,\delta (k), 
\end{equation}
where $f(k)$ is an any functions of $k$, which satisfies the condition $f (0) = 0$.
This gives
\begin{eqnarray}
 &\,&
 - \,i \,\langle \Psi^{(L_1)}_{n, k^\prime_x} \,\vert \, x \,\frac{\partial}{\partial y} \,
 \vert \,\Psi^{(L_1)}_{n, k_x} \rangle \nonumber \\
 &=& 
 \left\{ \frac{\partial}{\partial k_x} \,\frac{1}{l_B} \,
 \int_{- \,\infty}^\infty \,d \xi \,\psi_n (\xi^\prime) \,\frac{\partial}{\partial \xi} \,
 \psi_n (\xi) \right\}_{k^\prime_x = k_x} \times \,\,\delta (k^\prime_x - k_x) ,
\end{eqnarray}
if the following equality holds
\begin{equation}
 \int_{- \,\infty}^\infty \,d \xi \left. 
 \psi_n (\xi^\prime) \,\frac{\partial}{\partial \xi} \,\psi_n (\xi) \,
 \right\vert_{\xi^\prime = \xi} \ = \ 0,
\end{equation}
which can be easily verified to hold. We therefore find that
\begin{eqnarray}
 &\,&
 - \,i \,\langle \Psi^{(L_1)}_{n, k^\prime_x} \,\vert \,x \,\frac{\partial}{\partial y} \,
 \vert \,\Psi^{(L_1)}_{n, k_x} \rangle \nonumber \\
 &=&  
 \delta (k^\prime_x - k_x) \,\left( - \,l_B \,\frac{\partial}{\partial \xi} \right) \,
 \frac{1}{l_B} \,\int_{- \,\infty}^\infty \,d \xi \,\,\left.
 \psi_n (\xi^\prime) \,\frac{\partial}{\partial \xi} \,\psi (\xi) \,
 \right\vert_{\xi^\prime = \xi}
 \nonumber \\
 &=& - \,\delta (k^\prime_x - k_x) \,\int_{- \,\infty}^\infty \,
 \psi (\xi) \,\frac{\partial^2}{\partial \xi^2} \,\psi_n (\xi) 
 \ = \ \frac{1}{2} \,( 2 \,n + 1) \,\delta (k^\prime_x - k_x) . \ \ \ \ \ \ \ 
\end{eqnarray}
The second half part of (\ref{AppEq:L^can}) can be evaluated as
\begin{eqnarray}
 i \,\langle \Psi^{(L_1)}_{n, k^\prime_x} \,\vert \,y \,\frac{\partial}{\partial x} \,
 \vert \,\Psi^{(L_1)}_{n, k_x} \rangle \ &=& \ 
 - \,k_x \,\delta (k^\prime_x - k_x) \,\int_{- \,\infty}^\infty \,
 d y \,Y_n (y) \,y \,Y_n (y) \nonumber \\
 \ &=& \ - \,k_x \,\delta (k^\prime_x - k_x) \,y_0 \ = \ 
 - \,\frac{k_x^2}{e \,B} \,\delta (k^\prime_x - k_x) . \hspace{4mm} 
\end{eqnarray}
Combining the two terms, we therefore find that
\begin{equation}
 \langle \Psi^{(L_1)}_{n, k^\prime_x} \,\vert \, \hat{L}^{can}_z \,\vert \,
 \Psi^{(L_1)}_{n, k_x} \rangle \ = \ 
 \left\{\, n \, + \, \frac{1}{2} \ - \ \frac{k^2_x}{e \,B} \,\right\} \,
 \delta (k^\prime_x - k_x) .
\end{equation}

Since the calculation of the matrix element $\langle \Psi^{(L_1)}_{n, k^\prime_x} \,\vert \,
y^2 \,\vert \,\Psi^{(L_1)}_{n, k_x} \rangle$ is straightforward, here we show only
the answer given as
\begin{equation}
 \langle \Psi^{(L_1)}_{n, k^\prime_x} \,\vert \,
 y^2 \,\vert \,\Psi^{(L_1)}_{n, k_x} \rangle \ = \ 
 \frac{1}{e \,B} \,\left\{\, n \, + \, \frac{1}{2} \ + \ \frac{k^2_x}{e \,B} \right\} \,
 \delta (k^\prime_x - k_x) .
\end{equation}
Finally, as already pointed out, the calculation of the matrix element
$\langle \Psi^{(L_1)}_{n, k^\prime_x} \,\vert \, x^2 \,\vert \,\Psi^{(L_1)}_{n, k_x} \rangle$
is a little technical, so that here we show only the final answer by leaving the
explicit derivation to another Appendix B. The answer reads as
\begin{equation}
 \langle \Psi^{(L_1)}_{n, k^\prime_x} \,\vert \,x^2 \,\vert \,\Psi^{(L_1)}_{n, k_x} \rangle
 \ = \ \frac{1}{e \,B} \,\left( n + \frac{1}{2} \right) \,\delta (k^\prime_x - k_x)
 \ - \ \delta^{\prime \prime} (k^\prime_x - k_x) , \label{AppEq:x^2}
\end{equation}
where $\delta^{\prime \prime} (k)$ represents the second derivative of $\delta (k)$.
In this way, we now have all the necessary matrix elements as follows : 
\begin{eqnarray}
 \hspace{-8mm}
 \langle \Psi^{(L_1)}_{n, k^\prime_x} \,\vert \,\hat{L}^{can}_z \,\vert \,\Psi^{(L_1)}_{n, k_x} \rangle
 \ &=& \ \left\{\, n \, + \,\frac{1}{2} \,- \,\frac{k^2_x}{e \,B} \right\} \,
 \delta (k^\prime_x - k_x), \\
 \hspace{-8mm}
 e \,B \,\langle \Psi^{(L_1)}_{n, k^\prime_x} \,\vert \,y^2 \,\vert \,\Psi^{(L_1)}_{n, k_x} \rangle
 \ &=& \ \left\{\, n \, + \, \frac{1}{2} \,+ \,\frac{k^2_x}{e \,B} \right\} \,
 \delta (k^\prime_x - k_x), \\
 \hspace{-8mm}
 e \,B \,\langle \Psi^{(L_1)}_{n, k^\prime_x} \,\vert \,x^2 \,\vert \,\Psi^{(L_1)}_{n, k_x} \rangle
 \ &=& \ \left( n \, + \,\frac{1}{2} \right) \,\delta (k^\prime_x - k_x) \ - \
 e \,B \,\delta^{\prime \prime} (k^\prime_x - k_x) .
\end{eqnarray}
Using these answers, we can now readily write down the final answers for 
the matrix elements of the three OAM operators in the $\vert \,n, k_x \rangle$-basis.
The answers are summarized in Table 1 in the main text together with the matrix 
elements of the three momentum operators.

\section{Calculation of the matrix elements 
$\langle \Psi^{(L_1)}_{n, k^\prime_x} \,\vert\,x^2 \,\vert\,\Psi^{(L_1)}_{n, k_x} \rangle$}

\renewcommand{\theequation}{B.\arabic{equation} }
\setcounter{equation}{32}

The calculation starts with
\begin{eqnarray}
 &\,& \hspace{-6mm}
 \langle \Psi^{(L_1)}_{n, k^\prime_x} \,\vert\,x^2 \,\vert\,\Psi^{(L_1)}_{n, k_x} \rangle
 \nonumber \\
 \ &=& \ \left\{ \frac{1}{2 \,\pi} \,\int_{- \,\infty}^\infty \,d x \,
 e^{\,- \,i \,(k^\prime_x - k_x) \,x} \,x^2 \right\} \,
 \int_{- \,\infty}^\infty \,d y \,Y_n (y - y^\prime_0) \,Y_n (y - y_0) 
 \hspace{8mm} \nonumber \\
 \ &=& \ \left\{ - \,\frac{\partial^2}{\partial k^2_x} \,
 \delta (k^\prime_x - k_x) \right\} \,
 \int_{- \,\infty}^\infty \,d y \,Y_n (y - y^\prime_0) \,Y_n (y - y_0) .
\end{eqnarray}
With use of the following identity of Dirac's delta function, 
\begin{equation}
 \delta^{\prime \prime} (k) \,f (k) \ = \ f^{\prime \prime} (0) \,\delta (k)
 \ - \ 2 \,f^\prime (0) \,\delta^\prime (k) \ + \ 
 f (0) \,\delta^{\prime \prime} (k) ,
\end{equation}
we obtain
\begin{eqnarray}
 &\,& \hspace{-5mm}
 \langle \Psi^{(L_1)}_{n, k^\prime_x} \,\vert\,x^2 \,\vert\,\Psi^{(L_1)}_{n, k_x} \rangle
 \nonumber \\
 \ &=& \ - \,\left\{ \frac{\partial^2}{\partial k^2_x} \,
 \int_{- \,\infty}^\infty \,d y \,Y_n (y - y^\prime_0) \,
 Y_n (y - y_0) \right\}_{k^\prime_x = k_x} \,\,\delta (k^\prime_x - k_x) \nonumber \\
 \ &\,& \ + \ 2 \, \left\{ \frac{\partial}{\partial k_x} \,
 \int_{- \,\infty}^\infty \,d y \,Y_n (y - y^\prime_0) \,
 Y_n (y - y_0) \right\}_{k^\prime_x = k_x} \,\,
 \delta^\prime (k^\prime_x - k_x) \hspace{8mm} \nonumber \\
 \ &\,& - \ \left\{ \int_{- \,\infty}^\infty \,d y \,Y_n (y - y^\prime_0) \,
 Y_n (y - y_0) \right\}_{k^\prime_x = k_x} \,\,\delta^{\prime \prime} (k^\prime_x - k_x)
\end{eqnarray}
It is not so difficult to verify the equalities
\begin{equation}
 \left\{ \int_{- \,\infty}^\infty \,d y \,Y_n (y - y^\prime_0) \,
 Y_n (y - y_0) \right\}_{k^\prime_x = k_x} 
 \ = \ 1,
\end{equation}
and
\begin{equation}
 \left\{ \frac{\partial}{\partial k_x} \,
 \int_{- \,\infty}^\infty \,d y \,Y_n (y - y^\prime_0) \,
 Y_n (y - y_0) \right\}_{k^\prime_x = k_x} \ = \ 0 ,
\end{equation}
so that we show below how to evaluate
\begin{eqnarray}
 \left\{ \frac{\partial^2}{\partial k^2_x} \!\!
 \int_{- \,\infty}^\infty \!\! d y \,Y_n (y - y^\prime_0) \,
 Y_n (y - y_0) \right\}_{k^\prime_x = k_x} 
 \!\!\!\!\! &=& l^2_B \,\int_{- \,\infty}^\infty d \xi \,\psi_n (\xi) \,
 \frac{\partial^2}{\partial \xi^2} \,\psi_n (\xi). \hspace{8mm}
\end{eqnarray}
Using the familiar recursion formulas for the harmonic oscillator wave functions
\begin{eqnarray}
 \frac{\partial}{\partial \xi} \,\psi_n (\xi) \ &=& \ 
 \sqrt{\frac{n}{2}} \,\psi_{n-1} (\xi) \ - \ \sqrt{\frac{n+1}{2}} \,\psi_{n+1} (\xi), \\
 \frac{\partial^2}{\partial \xi^2} \,\psi_n (\xi) \ &=& \ 
 - \,\frac{2 \,n + 1}{2} \,\psi_n (\xi) \nonumber \\
 \ &+& \ \sqrt{\frac{n \,(n - 1)}{2}} \,\psi_{n-2} (\xi)
 \ + \ \sqrt{\frac{(n+1) \,(n+2)}{2}} \,\psi_{n+2} (\xi) , \ \ \ \ \ 
\end{eqnarray}
we find that
\begin{equation}
 \int_{- \,\infty}^\infty \,d \xi \,\psi_n (\xi) \,\frac{\partial^2}{\partial \xi^2} \,
 \psi_n (\xi) \ = \ - \,\frac{1}{2} \,(2 \,n + 1) ,
\end{equation}
which in turn gives
\begin{equation}
 \left\{ \frac{\partial^2}{\partial k^2_x} \,
 \int_{- \,\infty}^\infty \,d y \,Y_n (y - y^\prime_0) \,
 Y_n (y - y_0) \right\}_{k^\prime_x = k_x}
 \ = \ - \,l^2_B \,\,\frac{1}{2} \,(2 \,n + 1) .
\end{equation}
Collecting the above formulas, we finally get
\begin{equation}
\langle \Psi^{(L_1)}_{n, k^\prime_x} \,\vert\,x^2 \,\vert\,\Psi^{(L_1)}_{n, k_x} \rangle
 \ = \ \frac{1}{e \,B} \,\left( n + \frac{1}{2} \right) \,\delta (k^\prime_x - k_x)
 \ - \ \delta^{\prime \prime} (k^\prime_x - k_x) .
\end{equation}
which reproduces (\ref{AppEq:x^2}) in the Appendix A.

\section{Calculation of the matrix elements in the $\vert n, m \rangle$-basis}

\renewcommand{\theequation}{C.\arabic{equation} }
\setcounter{equation}{43}

Let us start with the calculation of the momentum operators. 
Using (\ref{Eq:p_cons_x_ladder}), we obtain
\begin{eqnarray}
 &\,&
 \langle \Psi^{(S)}_{n,m^\prime} \,\vert \,p^{cons}_x (\bm{A}^{(S)}) \,\vert \,\Psi^{(S)}_{n,m} \rangle
 \nonumber \\
 &=& - \,i \,\sqrt{\frac{e \,B}{2}} \,\,{}^A \langle n \,\vert  \,{}^B \langle n - m^\prime \,\vert \,
 b - b^\dagger \,\vert \,n \rangle^A \,\vert \,n - m \rangle^B \nonumber \\
 &=& - \,i \,\sqrt{\frac{e \,B}{2}} \left\{ \sqrt{n - m} \,\delta_{m^\prime, m+1}
 \, - \, \sqrt{n - m + 1} \,\delta_{m^\prime, m-1} \right\} . \ \ \ \ \ \ \ 
\end{eqnarray}
On the other hand, we get
\begin{eqnarray}
 &\,&
 \langle \Psi^{(S)}_{n,m^\prime} \,\vert \,p^{mech}_x (\bm{A}^{(S)}) \,\vert \,\Psi^{(S)}_{n,m} \rangle
 \nonumber \\
 &=& - \,i \,\sqrt{\frac{e \,B}{2}} \,\,{}^A \langle n \,\vert \,{}^B \langle n - m^\prime \,\vert \,
 a - a^\dagger \,\vert \,n \rangle^A \,\vert \,n - m \rangle^B \ = \ 0.
\end{eqnarray}
The matrix element of the canonical momentum operator can be calculated by using 
the relation $p^{can}_x = \frac{1}{2} \,\left( p^{mech}_x (\bm{A}^{(S)}) + 
p^{cons}_x (\bm{A}^{(S)}) \right)$, which gives
\begin{eqnarray}
 &\,&
 \langle \Psi^{(S)}_{n,m^\prime} \,\vert \,p^{can}_x  \,\vert \,\Psi^{(S)}_{n,m} \rangle
 \nonumber \\
 &=& - \,i \,\,\frac{1}{2} \,\sqrt{\frac{e \,B}{2}} \,\left\{ \sqrt{n - m} \,\delta_{m^\prime, m+1}
 \ - \ \sqrt{n - m + 1} \,\delta_{m^\prime, m-1} \right\} .
\end{eqnarray}
The matrix elements of the momentum operators between the eigen-states
$\vert \,\Psi^{(L_1)}_{n,m} \rangle$ are obtained by using the relation
$\vert \,\Psi^{(L_1)}_{n,m} \rangle = U \,\vert \,\Psi^{(S)}_{n,m} \rangle$.
For the canonical momentum operator, we get
\begin{eqnarray}
 \langle \Psi^{(L_1)}_{n,m^\prime} \,\vert \, p^{can}_x \,\vert \,\Psi^{(L_1)}_{n,m} \rangle \ &=& \ 
 \langle \Psi^{(S)}_{n,m^\prime} \,\vert \, U^\dagger \,p^{can}_x \,U \,\vert \,\Psi^{(S)}_{n,m} \rangle
 \nonumber \\
 \ &=& \ \langle \Psi^{(S)}_{n,m^\prime} \,\vert \, p^{can}_x \ + \ 
 \frac{1}{2} \,e \,B \,y \,\vert \,\Psi^{(S)}_{n,m} \rangle . \ \ \ \ \ \ \ 
\end{eqnarray}
The second part can be evaluated by using
\begin{equation}
 \frac{1}{2} \,e \,B \,y \ = \ i \,\,\frac{1}{2} \,\sqrt{\frac{e \,B}{2}} \,\,
 ( a \ - \ b \ - \ a^\dagger \ + \ b^\dagger ) ,
\end{equation}
which gives
\begin{eqnarray}
 &\,&
 \frac{1}{2} \,e \,B \,\langle \Psi^{(S)}_{n,m^\prime} \,\vert\,y \,\vert \,\Psi^{(S)}_{n,m} \rangle
 \nonumber \\
 &=&
 - \,i \,\,\frac{1}{2} \,\sqrt{\frac{e \,B}{2}} \,\left\{ \sqrt{n - m} \,\delta_{m^\prime, m+1}
 \, - \, \sqrt{n - m + 1} \,\delta_{m^\prime, m-1} \right\} .
\end{eqnarray}
Combining the two terms, we obtain
\begin{eqnarray}
 &\,&
 \langle \Psi^{(L_1)}_{n,m^\prime} \,\vert \, p^{can}_x \,\vert \,\Psi^{(L_1)}_{n,m} \rangle
 \nonumber \\
 &=& - \,i \,\sqrt{\frac{e \,B}{2}} \,\left\{ \sqrt{n - m} \,\delta_{m^\prime, m+1}
 \ - \ \sqrt{n - m + 1} \,\delta_{m^\prime, m-1} \right\} .
\end{eqnarray}
Similarly, we can readily show that
\begin{eqnarray}
 &\,&
 \langle \Psi^{(L_1)}_{n,m^\prime} \,\vert \, p^{cons}_x (\bm{A}^{(L_1)}) \,\vert \,\Psi^{(L_1)}_{n,m} \rangle
 \nonumber \\
 &=& - \,i \,\sqrt{\frac{e \,B}{2}} \,\left\{ \sqrt{n - m} \,\delta_{m^\prime, m+1}
 \, - \, \sqrt{n - m + 1} \,\delta_{m^\prime, m-1} \right\} , \ \ \ \ \ \ \ \ 
\end{eqnarray}
and that
\begin{equation}
 \langle \Psi^{(L_1)}_{n,m^\prime} \,\vert \, p^{mech}_x (\bm{A}^{(L_1)}) \,\vert \,\Psi^{(L_1)}_{n,m} \rangle
 \ = \ 0 .
\end{equation}

\vspace{2mm}
Next, we evaluate the matrix elements of the three OAM operators.
First, for the canonical OAM operator, we obtain
\begin{eqnarray}
 \langle \Psi^{(S)}_{n,m^\prime} \,\vert \,L^{can}_z \,\vert \,\Psi^{(S)}_{n,m} \rangle 
 &=& 
 {}^A \langle n \,\vert  \,{}^B \langle n - m^\prime \,\vert \, a^\dagger \,a - b^\dagger \,b \,
 \vert \,n \rangle^A \,\vert \,n - m \rangle^B \hspace{4mm} \nonumber \\
 &=&  n \,\delta_{m^\prime, m} \ - \ ( n - m) \,\delta_{m^\prime, m}
 \ = \ m \,\delta_{m^\prime, m} . \hspace{6mm}
\end{eqnarray}
Since $L^{cons}_z (\bm{A})$ reduces to $L^{can}_z$ in the
symmetric gauge, we naturally get
\begin{equation}
 \langle \Psi^{(S)}_{n,m^\prime} \,\vert \,L^{cons}_z (\bm{A}^{(S)}) \,\vert \,\Psi^{(S)}_{n,m} \rangle 
 \ = \ m \,\delta_{m^\prime, m} ,
\end{equation}
Finally, for the matrix element of the mechanical OAM operator, we find that
\begin{equation}
 \langle \Psi^{(S)}_{n,m^\prime} \,\vert \,L^{mech}_z (\bm{A}^{(S)}) \,\vert \,\Psi^{(S)}_{n,m} \rangle 
 \ = \ ( 2 \,n + 1) \,\delta_{m^\prime, m} ,
\end{equation}

Next, we evaluate the matrix elements of the OAM operators between the
eigen-states $\vert \,\Psi^{(L_1)}_{n,m} \rangle$, which also belong to the 
$\vert n,m \rangle$-basis class.
Here we start with the matrix elements of the canonical OAM operator
given by
\begin{eqnarray}
 \langle \Psi^{(L_1)}_{n,m^\prime} \,\vert \,L^{can}_z \,\vert \,\Psi^{(L_1)}_{n,m} \rangle \ &=& \ 
 \langle \Psi^{(S)}_{n,m^\prime} \,\vert \,U^\dagger \,L^{can}_z \,U \,\vert \,\Psi^{(S)}_{n,m} \rangle
 \nonumber \\
 \ &=& \ \langle \Psi^{(S)}_{n,m^\prime} \,\vert \, L^{can}_z \ + \ 
 \frac{1}{2} \,e \,B \,(x^2 - y^2) \,\vert \,\Psi^{(S)}_{n,m} \rangle . \hspace{4mm}
\end{eqnarray}
The second part can be calculated by using the relations
\begin{eqnarray}
 x^2 \ &=& \ l^4_B \,(\,\Pi_y \ - \ \tilde{\Pi}_y )^2 \ = \ 
 \frac{1}{2} \,\,l^2 _B \,(\,a \ + \ a^\dagger \ + \ b \ + \ b^\dagger )^2,
\end{eqnarray}
and
\begin{eqnarray}
 y^2 \ &=& \ l^4_B \,(\,\Pi_x \ - \ \tilde{\Pi}_x )^2 \ = \ 
 - \,\frac{1}{2} \,l^2_B \,(\, a \ - \ a^\dagger \ - \ b \ + \ b^\dagger )^2,
\end{eqnarray} 
which gives
\begin{eqnarray}
 &\,&
 x^2 \ - \ y^2 \nonumber \\
 &\,& \hspace{0mm} \ = \ l^2_B \,\,
 \left\{ \,a^2 \,+ \,b^2 \,+ \,(a^\dagger)^2 \,+ \,(b^\dagger)^2 
 \,+ \,a \,b^\dagger \,+ \,b \,a^\dagger
 \,+ \,b^\dagger \,a \,+ \,a^\dagger \,b \, \right\} .\hspace{6mm}
\end{eqnarray}
After some tedious but straightforward algebra, we find that
\begin{eqnarray}
 \frac{1}{2} \,e \,B \,\langle \Psi^{(S)}_{n,m^\prime} \,\vert \,x^2 - y^2 \,
 \vert \,\Psi^{(S)}_{n,m} \rangle
 &=& \frac{1}{2} \,\left\{ \sqrt{(n-m) \,(n-m+1)} \,\delta_{m^\prime, m+2}
 \right. \nonumber \\
 &+& \left. \sqrt{(n-m+1) \,(n-m+2)} \,\delta_{m^\prime,m-2} \right\} , \hspace{8mm}
\end{eqnarray}
which in turn gives
\begin{eqnarray}
 \langle \Psi^{(L_1)}_{n,m^\prime} \,\vert \,L^{can}_z \,\vert \,\Psi^{(L_1)}_{n,m} \rangle 
 &=& m \,\delta_{m^\prime,m} \nonumber \\
 &+& 
 \frac{1}{2} \,\left\{ \sqrt{(n-m) \,(n-m+1)} \,\delta_{m^\prime, m+2} \right.
 \nonumber \\
 &+& \left. 
 \sqrt{(n-m+1) \,(n-m+2)} \,\delta_{m^\prime,m-2} \right\} .  
\end{eqnarray}
Since both of $L^{cons}_z (\bm{A})$ and $L^{mech}_z (\bm{A})$ transform
covariantly under a gauge transformation, their matrix elements between
the eigen-state $\vert \,\Psi^{(L_1)}_{n,m} \rangle$ can easily be obtained from
those between the eigen-states $\vert \,\Psi^{(S)}_{n,m} \rangle$ as
\begin{eqnarray}
 \langle \Psi^{(L_1)}_{n,m^\prime} \,\vert \,L^{cons}_z (\bm{A}^{(L_1)}) \,\vert \,\Psi^{(L_1)}_{n,m} \rangle
 &=& \langle \Psi^{(S)}_{n,m^\prime} \,\vert \,L^{cons}_z (\bm{A}^{(S)}) \,\vert \,\Psi^{(S)}_{n,m} \rangle
 \nonumber \\
 &=& m \,\delta_{m^\prime, m} ,
\end{eqnarray}
and
\begin{eqnarray}
 \langle \Psi^{(L_1)}_{n,m^\prime} \,\vert \,L^{mech}_z (\bm{A}^{(L_1)}) \,\vert \,\Psi^{(L_1)} \rangle
 &=& \langle \Psi^{(S)}_{n,m^\prime} \,\vert \,L^{mech}_z (\bm{A}^{(S)}) \,\vert \,\Psi^{(S)}_{n,m} \rangle
 \nonumber \\
 &=& ( 2 \,n + 1 ) \,\delta_{m^\prime, m} .
\end{eqnarray}
%




\begin{thebibliography}{}

\bibitem{Landau1930}
Landau, L. D. :
Diamagnetisms der Metalle,
Z. Phys. {\bf 64}, 629-637 (1930)

\bibitem{Landau-Lifschitz}
Landau, L. D., Lifshitz, E. M. :
Quantum Mechanics : Non-Relativistic Theory,
Course of Theoretical Physics, 3rd ed., Pergamon, New York (1997)

\bibitem{Vallentine1998}
Vallentine, L. E. : 
Quantum Mechanics : A Modern Development, 
World Scientific Publishing Co. Pte. Ltd. (1998)

\bibitem{VGW2006}
Vagner, I. D., Gvozdikov, V. M., Wyder, P. : 
Quantum mechanics of electrons in strong magnetic field,
HIT Journal of Science and Engineering, Vol.{\bf 3} Issue {\bf 1} 5-55 (2006)

\bibitem{Goerbig2009}
Goerbig, M. O. : 
Quantum Hall Effects, {\it arXiv:0909.1998} (2009)

\bibitem{Tong2016}
Tong, D. : 
Quantum Hall Effects, {\it arXiv:1606.06687} (2016)

\bibitem{Murayama}
Murayama, H. : Landau levels, 
https://hitoshi.berkley.edu/221a/landau.pdf

\bibitem{Yoshioka2002}
Yoshioka, D. :  
The Quantum Hall Effects, Springer (2002) pp.20-23

\bibitem{Konstantinou2016}
Konstantinou, G., Moulopoulos, K. : 
Generators of dynamical symmetries and the correct gauge transformation
in the Landau level problem : use of pseudomomentum and pseudo-angular
momentum,
Eur. J. Phys. {\bf 37}, 065401/1-15 (2016)

\bibitem{Konstantinou2017}
Konstantinou, G., Moulopoulos, K. : 
The ``forgotten'' pseudomomenta and gauge changes in generalized Landau
level problems : spatially nonuniform magnetic and temporally varying
electric fields,
Int. J. Theor. Phys. {\bf 56} 1484-1503 (2017)

\bibitem{WKZ2018}
Wakamatsu, M., Kitadono, Y., Zhang, P. M. :  
The issue of gauge choice in the Landau problem and the physics of
canonical and mechanical orbital angular momentum,
Ann. Phys. {\bf 392} 287-322 (2018)

\bibitem{DeWitt1962}
DeWitt, B. S. : 
Quantum Theory without Electromagnetic Potentials,
Phys. Rev. {\bf 125} 2189-2191 (1962)

\bibitem{Chen2008}
Chen, X. S., L\"{u}, X. F., Sun, W. M., Wang, F., Goldman, T. :  
Spin and Orbital Angular Momentum in Gauge Theories: Nucleon Spin Structure
and Multipole Radiation Revisited,
Phys. Rev. Lett. {\bf 100} 232002/1-4 (2008)

\bibitem{Chen2009}
Chen, X. S., Sun, W. M., L\"{u}, X. F., Wang, F. , Goldman, T. :  
Do Gluons Carry Half of the Nucleon Momentum?,
Phys. Rev. Lett. {\bf 103} 062001/1-4 (2009)

\bibitem{Hatta2011}
Hatta, Y. :  
Gluon polarization in the nucleon demystified
Phys. Rev. {\bf D 84} 04701(R)/1-4 (2011)

\bibitem{Lorce2013}
Lorc\'{e}, C. :  
Gauge-covariant canonical formalism revisited with application to the
proton spin decomposition,
Phys. Rev. {\bf D 88} 044037/1-8 (2013)

\bibitem{Review_LL2014}
Leader, E., Lorc\'{e}, C. :  
The angular momentum controversy : What's it all about and does it matter?,
Phys. Rep. {\bf 541}, 163-248 (2014)

\bibitem{Review_Waka2014}
Wakamatsu, M. : 
Is gauge-invariant complete decomposition of the nucleon spin possible?,
Int. J. Mod. Phys. {\bf A 29} 1430012/1-52 (2014)

\bibitem{Haugset1993}
Haugset, T., Ruud, J. Aa., Ravndal, F. :  
Gauge invariance of Landau levels,
Physica Scripta {\bf 47} 715-719 (1993)

\bibitem{Govaerts2009}
Govaerts, J., Hounkonnou, M. N., Mweene, H. V. : 
Variations on the planar Landau problem : canonical transformations,
a purely linear potential and the half-plane,
J. Phys. A : Math. Theor. {\bf 42} 485209/1-19 (2009) 

\bibitem{WKZZ2020}
Wakamatsu, M., Kitadono, Y., Zou, L., Zhang, P. :  
The physics of helical electron beam in a uniform magnetic field
as a testing ground of gauge principle,
Phys. Lett. {\bf A384} 126415/1-7 (2020)

\bibitem{WKZZ2021}
Wakamatsu, M., Kitadono, Y., Zou, L., Zhang, P. : 
Revisiting the compatibility problem between the gauge principle
and the observability of the canonical orbital angular momentum in
the Landau problem, 
Ann. Phys. {\bf 434} 168647/1-26 (2021)

\bibitem{Cohen-Tannoudji1989}
Cohen-Tannoudji, C., Dupont-Roc, J.,  Grynberg, G. :  
Photons and Atoms - Introduction to Quantum Electrodynamics,
John Wiley $\&$ Sons, Inc. (1989)

\bibitem{Zangwill2013}
Zangwill, A. :  
Electrodynamics, 
Cambridge University Press, United Kingdam (2013)

\bibitem{JM1990}
Jaffe, R. L.,  Manohar, A. : 
The $g_1$ problem : Deep inelastic electron scattering and the spin
of the nucleon,
Nucl. Phys. {\bf B 337} 509-546 (1990)

\bibitem{Ji1997}
Ji, X. :  
Gauge-Invariant Decomposition of Nucleon Spin,
Phys. Rev. Lett. {\bf 78} 610-613 (1997)

\bibitem{Waka2010}
Wakamatsu, M. :  
Gauge-invariant decomposition of nucleon spin,
Phys. Rev. {\bf D 81} 114010/1-9 (2010)

\bibitem{Waka2011}
Wakamatsu, M. :  
Gauge- and frame-independent decomposition of nucleon spin,
Phys. Rev. {\bf D 83} 014012/1-16 (2011)

\bibitem{Zee2010}
Zee, A. : 
Quantum Field Theory in a Nutshell,
Princeton University Press, USA (2010)

\bibitem{Schwartz2014}
Schwartz, M. : 
Quantum Field Theory and the Standard Model,
Cambridge University Press, United Kingdam (2014)



\end{thebibliography}




\end{document}